\documentclass[11pt]{article}

\usepackage[totalheight = 23cm, totalwidth = 17cm]{geometry}
\usepackage{amssymb,amsmath,amsfonts,amsbsy,graphicx}

\def\mathbi#1{\textbf{\em #1}}

\newcommand{\mpl}{m_{\rm Pl}}

\newcommand{\calD}{{\cal D}}
\newcommand{\calG}{{\cal G}}
\newcommand{\calH}{{\cal H}}
\newcommand{\calO}{{\cal O}}
\newcommand{\calP}{{\cal P}}
\newcommand{\calR}{{\cal R}}

\renewcommand{\theequation}{\arabic{section}.\arabic{equation}}

\newcommand{\dis}[1]{\begin{equation}\begin{split}#1\end{split}\end{equation}}

\begin{document}

\begin{titlepage}

\rightline{\footnotesize{APCTP-Pre2016-002}} \vspace{-0.2cm}
\rightline{\footnotesize{CTPU-16-03}} \vspace{-0.2cm}
\rightline{\footnotesize{MAD-TH-16-01}} \vspace{-0.2cm}

\begin{center}

\vskip 1.0 cm

{\LARGE \bf Path integral for multi-field inflation}

\vskip 1.0cm

{\large
Jinn-Ouk Gong$^{a,b}$, Min-Seok Seo$^{c}$ and Gary Shiu$^{d,e}$
}

\vskip 0.5cm

{\it
$^{a}$Asia Pacific Center for Theoretical Physics, Pohang 37673, Korea
\\
$^{b}$Department of Physics, Postech, Pohang 37673, Korea
\\
$^{c}$Center for Theoretical Physics of the Universe, 
\\
Institute for Basic Science, 34051 Daejeon, Korea
\\
$^{d}$Department of Physics, University of Wisconsin-Madison, Madison, WI 53706, USA
\\
$^{e}$Department of Physics \& Institute for Advanced Study,
\\
Hong Kong University of Science and Technology, Clear Water Bay, Hong Kong
}

\vskip 1.2cm

\end{center}

\begin{abstract}

We develop the path integral formalism for studying cosmological perturbations in multi-field inflation, which is particularly well suited to study quantum theories with gauge symmetries such as diffeomorphism invariance. We formulate the gauge fixing conditions based on the Poisson brackets of the constraints, from which we derive two convenient gauges that are appropriate for multi-field inflation. We then adopt the in-in formalism to derive the most general expression for the power spectrum of the curvature perturbation including the corrections from the interactions of the curvature mode with other light degrees of freedom. We also discuss the contributions of the interactions to the bispectrum.

\end{abstract}

\end{titlepage}

\newpage

\section{Introduction}
\setcounter{equation}{0}

Inflation~\cite{inflation}, an early period of accelerated expansion, is the leading paradigm describing our primordial universe for several reasons. First of all, it resolves the initial condition problems behind the homogeneity, isotropy, and flatness of the observed universe~\cite{books}, which has been well confirmed through the observations of the cosmic microwave background (CMB). Moreover, it explains how small quantum fluctuations are stretched beyond the horizon to become the seeds for large scale inhomogeneities~\cite{books,slava}. The properties of these primordial fluctuations, and in turn the underlying physics, can be studied by scrutinizing cosmological observables in the CMB and large scale structure data. Indeed, most recent observations on the CMB have placed stringent constraints on models of inflation and properties of primordial perturbations produced during inflation~\cite{planck}.

Quantum field theoretical approaches are essential in describing the primordial fluctuations. Two prevailing frameworks to describe quantum fields are the operator formalism and the path integral approach. Whereas the two approaches give the same physics, there are several virtues of the path integral approach over the operator formalism. On the one hand, a theory with gauge symmetries can be readily quantized in the path integral formalism. Since gauge fields constitute a constrained system, the underlying gauge symmetry is interpreted as a redundancy of the physical degrees of freedom. A systematic way of quantizing such a gauged system is well developed in the path integral formulation~\cite{gauge-pathint}. Inflationary cosmology concerns with the behavior of scalar fields under gravity, which has an invariance under general coordinate transformation (or diffeomorphism) as a gauge symmetry~\cite{Hidaka:2014fra}. In this regard, the path integral formalism is well suited to study the quantum behavior in inflationary cosmology. On the other hand, the path integral approach reveals more than what is obtained from perturbation theory. Non-perturbative effects such as instantons, and the relations between Green's functions resulting from the underlying symmetries of the theory are more manifest in terms of the path integral. For example, the consistency relation in the squeezed limit can be extended to the Slavnov-Taylor identities of spatial diffeomorphism~\cite{consistency}.

The first virtue of the path integral formalism in inflation was pioneered in~\cite{Anderegg:1994xq} in terms of the linear gauge-invariant perturbation [see our eq. \eqref{Eq:multiredef}] and extended to the non-linear regime in the context of single-field inflation  in~\cite{Prokopec:2010be}. Meanwhile, it is possible that inflation takes place in the presence of interactions between the  ``inflaton'' and other degrees of freedom, and the classical trajectory of the inflaton should be determined in a multi-dimensional field space. Regarding quantum fluctuations around the classical solutions, there exists one component which transforms non-linearly under time translation. This corresponds to the Goldstone mode, since the Friedmann-Robertson-Walker (FRW) spacetime deviating from perfect de Sitter breaks the time translational invariance spontaneously~\cite{Cheung:2007st}. As a result, during inflation, with the slow-roll parameter $\epsilon \equiv -\dot{H}/H^2$ satisfying $1\gg \epsilon \gg \dot{\epsilon}/H$, the massless Goldstone mode decouples from gravity according to the Goldstone boson equivalence theorem~\cite{equi-theorem} for energy scales larger than $\sqrt{\epsilon}H$. This Goldstone mode corresponds to the quantum fluctuation of a gauge-invariant scalar degree of freedom which leads to the curvature perturbation (see Section~\ref{subsec:quadratic}), and its $n$-point correlation functions are key observables of the early universe. Effects of other fields which distinguish the multi-field model from single-field inflation could show up in the correction functions of the Goldstone mode and can thus be tested in future observations. The effective field theory of inflation~\cite{Cheung:2007st,Weinberg:2008hq} has been generalized to multi-field case~\cite{Senatore:2010wk}. However, imposing shift symmetry on all the scalar fields and decoupling gravity, which meanwhile greatly simplifies the analysis for the effective action~\cite{Senatore:2010wk}, also forbids interaction terms that give rise to interesting and important multi-field dynamics, such as the mixings arising from sharp turns in field space (see e.g.~\cite{Shiu:2011qw}). This motivates our study of multi-field inflation in the path integral formalism.
In this work, we go beyond the pioneering works~\cite{Anderegg:1994xq,Prokopec:2010be} and further develop the path integral approach for  multi-field inflation. In generalizing these works, we adopt the methods and notation developed in~\cite{Prokopec:2010be}. Other than formulating the gauge-fxiing conditions for multi-field inflation, we identify the Goldstone mode and its interactions with other degrees of freedom. Our findings enable us to derive the multi-field effects on the power spectrum and non-Gaussianity of curvature perturbation.

This article is organized as follows. In Section~\ref{Sec:pathconstraints}, we briefly review the quantization of constrained system in the path integral formalism, bearing in mind its applications to systems with gauge symmetry. The action for systems with  gravity and scalar fields is written and modified to an appropriate form for quantization in Section~\ref{Sec:haimiltonian}. Especially for quantization of gravity, the Arnowitt-Deser-Misner (ADM) decomposition~\cite{Arnowitt:1962hi} is used, as the nature of gravity as a constrained system becomes evident. In Section~\ref{Sec:quantization}, the action of quantum fluctuations around the classical solutions is provided. After extracting the Goldstone mode, the quadratic actions for cosmological perturbations are given. We also revisited the gauge fixing condition for path integral quantization. In Section~\ref{Sec:2point}, we adopt the in-in formalism to study the two-point correlation functions. We demonstrate here how the power spectrum of the Goldstone mode is corrected due to the quadratic mixing with other degrees of freedom. In Section~\ref{Sec:higherorder} a further study of the higher-order correlation functions is made, using the three-point correlator as an example. Finally, we conclude in Section~\ref{Sec:conclusion}.

\section{Path integral for constrained system}
\label{Sec:pathconstraints}
\setcounter{equation}{0}

Gravity has a gauge symmetry, an invariance under  general coordinate transformation or diffeomorphism, and its quantization is made most explicitly consistent in the path integral formalism. In general, the action with gauge symmetry contains both physical and unphysical degrees of freedom and gauge invariance is interpreted as a redundancy of the physical degrees of freedom. Some of unphysical degrees of freedom play a role of Lagrange multipliers for constraints. As constraints hold over every time slice, they become conserved quantities, or generators of gauge transformation.

To see the situation in detail \cite{Gaugenature}, consider a Hamiltonian system defined on a 2$f$-dimensional phase space $(q_1,\cdots, q_f;p_1,\cdots, p_f)$ consisting of  both physical and unphysical degrees of freedom,
 \dis{H_T=h(q_1,\cdots, q_f;p_1,\cdots, p_f)+\sum_{m=1}^r \lambda_m \chi_m(q_1,\cdots, q_f;p_1,\cdots, p_f) \, ,}
where Lagrange multipliers $\lambda_m$ being combinations of unphysical degrees of freedom, and constraints $\chi_m$ are interpreted as generators of gauge symmetry (first class constraints), forming a closed algebra. Due to the $r$ constraints $\chi_m$ ($m=1,\cdots, r$), the number of degrees of freedom diminishes to ($2f-r$), but we still have unphysical degrees of freedom because the final phase space of the physical degrees of freedom should always form even-dimensional phase space: eventually $(2f-2r)$ physical degrees of freedom should remain. Actually, on the hypersurface $\chi_m=0$ defined in the phase space, unphysical degrees of freedom are not completely eliminated. The phase space on $\chi_m=0$ is regarded as copies (called `orbits') of physical degrees of freedom and each copy is labelled by different values of $\lambda_m$. This calls for gauge fixing by imposing $r$ more conditions $\psi_m (q_1,\cdots, q_f;p_1,\cdots, p_f)=0$. These gauge fixing conditions satisfy the Faddeev-Popov determinant det$([\chi_m, \psi_n])\ne 0$ such that just one orbit is chosen. Recall that $\chi_m$ are the generators of gauge transformation. If det$([\chi_m, \psi_n])= 0$, the functions $\psi_n$ are gauge invariant, so they do not choose just one orbit, but a set of orbits related by gauge transformation. By choosing gauge fixing conditions satisfying $[\psi_m, \psi_n]=0$, through the canonical transformation, we can identify $\psi_m$ as unphysical canonical momenta $p_m$ ($m=1, \cdots, r$), then the remaining $(f-r)$ momenta $(p^*_1, \cdots p^*_{f-r})$ are chosen to be physical. Since det$([\chi_m, \psi_n])={\rm det}([\chi_m, p_n])={\rm det}(i\partial \chi_m/\partial q_n) \ne 0$, it is guaranteed that we obtain $q_m=f_m(q^*_i, p^*_i)$ ($m=1, \cdots, r$) by inverting $\chi_m(q_1,\cdots, q_f;p^*_1,\cdots, p^*_{f-r};p_1=\cdots = p_r=0)=0$. As a result, there remain $2(f-r)$ physical, constrained variables $(q^*_1,\cdots, q^*_{f-r};p^*_1,\cdots, p^*_{f-r})$, which are quantized in a normal way, $[q^*_i, p^*_j]=i$. They live on the intersection of $2r$ hypersurfaces $\chi_m=\psi_m=0$ and their dynamics are regulated by the constrained Hamiltonian defined on this intersection,
 \dis{H^*(q^*_i, p^*_i)=h(q_1,\cdots, q_f;p_1,\cdots, p_f) |_{p_m=0, q_m=f_m(q^*_i, p^*_i)} \, .}

In the path integral formalism, such a procedure corresponds to putting the Faddeev-Popov determinant ${\rm det}([\chi_m, \psi_n])$ to the propagation kernel written in terms of both physical and unphysical variables. To see this, we begin with the propagation kernel with {\em physical} variables, 
\dis{K(t_f;t_i)=\int \prod_{i=1}^{f-r} {\cal D}q^*_i {\cal D}p^*_i e^{i \int_{t_i}^{t_f} dt [p^*_i q^*_i-H^*(q^*_i, p^*_i)]} \, .}
Since 
\dis{\prod_{m=1}^r \delta(q_m-f_m) = \prod_{m=1}^r \delta(\chi_m)\frac{\partial(\chi_1,\cdots, \chi_r)}{\partial(q_1,\cdots, q_r)} =\prod_{m=1}^r \delta(\chi_m){\rm det}([\chi_m, \psi_n]),}
the integration measure is rewritten as
\dis{\prod_{i=1}^{f-r} {\cal D}q^*_i {\cal D}p^*_i &=\prod_{i=1}^{f-r} {\cal D}q^*_i {\cal D}p^*_i \prod_{m=1}^r {\cal D}q_m {\cal D}p_m \delta(p_m)\delta(q_m-f_m(p^*_i, q^*_i))
\\
&=\prod_{i=1}^{f} {\cal D}q_i {\cal D}p_i \prod_{m=1}^r \delta(\psi_m)\delta(q_m-f_m)
\\
&=\prod_{i=1}^{f} {\cal D}q_i {\cal D}p_i \prod_{m=1}^r \delta(\psi_m)\delta(\chi_m){\rm det}([\chi_m, \psi_n]) \, .}
From this, the propagation kernel is expressed in terms of both physical and {\em unphysical} variables with the unconstrained Hamiltonian $H_T$:
\dis{K(t_f;t_i)&=\int \prod_{i=1}^{f} {\cal D}q_i {\cal D}p_i \prod_{m=1}^r \delta(\psi_m)\delta(\chi_m){\rm det}([\chi_m, \psi_n])e^{i \int_{t_i}^{t_f}dt[ p_i q_i-H^*(q^*_i, p^*_i)]}
\\
&=\int \prod_{i=1}^{f} {\cal D}q_i {\cal D}p_i \prod_{m=1}^r {\cal D}\lambda_m \delta(\chi_m){\rm det}([\chi_m, \psi_n])e^{i \int_{t_i}^{t_f}dt [p_i q_i-H^*(q^*_i, p^*_i)-\lambda_m \chi_m]}
\\
&=\int \prod_{i=1}^{f} {\cal D}q_i {\cal D}p_i \prod_{m=1}^r {\cal D}\lambda_m \delta(\chi_m){\rm det}([\chi_m, \psi_n])e^{i \int_{t_i}^{t_f} dt [p_i q_i-H_T]} \, . 
\label{Eq:pathint}}

In the quantization of system including gauge fields like gravity, the path integral quantization proceeds in the reversed direction:
\begin{itemize}

 \item[1] Rewrite the Lagrangian in terms of the canonical variables, Lagrangian multipliers, constraints, and unconstrained Hamiltonian $H_T$ (For gravity, this corresponds to ADM decomposition). The constraints $\chi_m$ form closed algebra and they are interpreted as generators of gauge transformation.
  
 \item[2] Impose appropriate gauge fixing conditions and insert the Faddeev-Popov determinant to obtain the last of \eqref{Eq:pathint}. The Faddeev-Popov determinant appears in the Feynman rule by introducing ghost fields $c^m$ and $\bar{c}^m$. Then Lagrangian has additional term, $\bar{c}^m[\chi_m, \psi_n]c^n$.

\end{itemize}

\section{Hamiltonian formalism}
\label{Sec:haimiltonian}
\setcounter{equation}{0}

\subsection{First-order form}

Consider a multi-field system coupled to the Einstein gravity: 
\begin{equation}
S = \int d^4x \sqrt{-g} \left[ \frac{\mpl^2}{2}R - \frac{1}{2}{\cal G}_{ab}g^{\mu\nu}\partial_\mu\phi^a\partial_\nu\phi^b - V(\phi^a) \right] \, ,
\end{equation}
where $\calG_{ab}$ is the metric for $N$-dimensional field space ($a, b = 1, \cdots, N$). For quantization of gravity, we begin with the ADM decomposition,
\dis{ds^2=-N^2 dt^2+\gamma_{ij}(N^i dt + dx^i)(N^j dt + dx^j) \, .}
In terms of the extrinsic curvature
\begin{equation}
K_{ij} = \frac{1}{2N} \left( \partial_t\gamma_{ij} - D_iN_j - D_jN_i \right) \, ,
\end{equation}
with $D_i$ being a covariant derivative with respect to $\gamma_{ij}$, the pure gravity part of the action can be written as 
\begin{equation}
S_G = \int d^4x \sqrt{-g} \frac{\mpl^2}{2}R = \int d^4x \sqrt{\gamma}N \frac{\mpl^2}{2} \left( R^{(3)}+K_{ij}K^{ij}-K^2 \right) \, ,
\end{equation}
where $R^{(3)}$ is the curvature of three-dimensional hypersurface of constant $t$, constructed from $\gamma_{ij}$. From this form of the action, the canonical momentum of $\gamma_{ij}$ is obtained as 
\begin{equation}
\Pi^{ij} = \frac{\delta S_G}{\delta \partial_t \gamma_{ij}}=\frac{\mpl^2}{2}\sqrt{\gamma}(K^{ij}-\gamma^{ij}K) \, .
\end{equation}
Therefore, the pure gravity part of the action is written in the first-order form,
\begin{equation}
S_G = \int d^4x \left( \Pi^{ij}\partial_t{\gamma}_{ij} -N{\cal H}_G-N_i {\cal H}_G^i \right) \, ,
\end{equation}
where
\begin{align}
\calH_G & = \frac{2}{\mpl^2\sqrt{\gamma}} \left( \Pi^{ij} \Pi_{ij} - \frac12 \Pi^2 \right) - \frac{\mpl^2}{2}\sqrt{\gamma} R^{(3)},
\\
\calH_G^i & = -2 \left( \partial_j\Pi^{ij} + \Gamma^i_{jk}\Pi^{jk} \right) \, ,
\end{align}
with $\Pi \equiv \Pi^i{}_i$. Here, the lapse $N$ and the shifts $N_i$ play the role of Lagrangian multipliers for the constraints ${\cal H}$ and ${\cal H}^i$ respectively, which are interpreted as the generators of diffeomorphism. This will become clear in Section~\ref{subsec:poisson}.

Likewise, we can write the matter part of the action in the first-order form as
\begin{equation}
S_M = \int d^4x \left( \Pi_a\partial_t{\phi}^a - N\calH_M - N_i\calH_M^i \right) \, ,
\end{equation}
where
\begin{align}
\Pi_a & = \frac{\sqrt{\gamma}}{N}\calG_{ab} \left( \partial_t\phi^b - N^i\partial_i\phi^b \right) \, ,
\\
{\cal H}_M & = \frac{{\cal G}^{ab}}{2\sqrt{\gamma}}\Pi_a\Pi_b+\frac{1}{2}\sqrt{\gamma}\gamma^{ij}{\cal G}_{ab}\partial_i\phi^a \partial_j \phi^b +\sqrt{\gamma} V(\phi^a) \, ,
 \\
{\cal H}_M^i & = \Pi_a \partial^i \phi^a \, .
\end{align}
In summary, the action in the first-order form is given by
\begin{equation}
S = \int d^4x \left[ \Pi^{ij}\partial_t{\gamma}_{ij} + \Pi_a\partial_t{\phi}^a - N\left(\calH_G+\calH_M \right) - N_i \left( \calH_G^i + \calH_M^i \right) \right] \, .
\end{equation}
In view of diffeomorphism, $\Pi^{ij}/\sqrt{\gamma}$ and $\Pi_a/\sqrt{\gamma}$ behave as tensors. Unphysical $N$ and $N_i$ become Lagrange multipliers accompanying constraints $\calH_G+\calH_M$ and $\calH_G^i+\calH_M^i$, respectively.

We are interested in the cosmological perturbations around the classical background. First, the graviton corresponds to the quantum fluctuation around the FRW background:
\begin{align}
\Pi^{ij} & = \frac{P(t)}{6a(t)} \left[ \delta^{ij}+\pi^{ij}(t,\mathbi{x}) \right] \, ,
\\
\gamma_{ij} & = a^2(t) \left[ \delta_{ij}+h_{ij}(t,\mathbi{x}) \right] \, ,
\\
N & = \bar{N}(t)+n(t,\mathbi{x}) \, ,
\\
N^i & = N^i(t,\mathbi{x}) \, .
\end{align}
Note that for ${\bar N}=1$, $t$ is just the cosmological time, whereas for ${\bar N}=a(t)$, $t$ is the conformal time. The quantum fluctuations around the classical backgrounds $\phi^a_0(t)$ are written as
\begin{align}
\phi^a & = \phi_0^a(t) + \varphi^a(t,\mathbi{x}) \, ,
\\
\Pi_a & = P_a(t) + \pi_a(t,\mathbi{x}) \, .
\end{align}
Classical solutions are obtained from the zeroth order action:
\begin{equation}
S_0 = \int d^4x \left[ P\partial_t a + P_a\partial_t\phi_0^a - {\bar N} \left( -\frac{P^2}{12\mpl^2 a}+\frac{\calG^{ab}P_aP_b}{2a^3}+a^3V \right) \right] \, .
\end{equation}
Varying this with respect to the classical backgrounds, we obtain the equations of motion:
\begin{alignat}{4}
\frac{\delta S_0}{\delta P} = 0 & : & \qquad \dot{a} & = -\frac{P}{6\mpl^2a} \, ,
\\
\frac{\delta S_0}{\delta a} = 0 & : & \qquad \dot{P} & = -\frac{P^2}{12\mpl^2a^2} + \frac{3\calG^{ab}P_aP_b}{2a^4} - 3a^2V \, ,
\\
\frac{\delta S_0}{\delta P_a} = 0 & : & \qquad \dot\phi_0^a & = \frac{\calG^{ab}P_b}{a^3} \, ,
\\
\frac{\delta S_0}{\delta\phi_0^a} = 0 & : & \qquad \dot{P}_a & = -a^3V_a - \frac{1}{2a^3} \calG^{cd}{}_{,a} P_cP_d \, ,
\\
\frac{\delta S_0}{\delta\bar{N}} = 0 & : & \qquad \frac{P^2}{12\mpl^2a} & = \frac{\calG^{ab}P_aP_b}{2a^3} + a^3V \, ,
\end{alignat}  
where ${\dot X}\equiv dX/{\bar N}dt$ and $V_a \equiv \partial V/\partial \phi^a$. From here on, we denote an overdot as a derivative with respect to the cosmic time $t$. But we will leave $\bar{N}$ in the action, so that changing to the conformal time is more convenient. Combining these equations, we obtain the following familiar relations:
\begin{align}
\label{eq:friedmann}
& H^2 = \frac{1}{3\mpl^2} \left(\frac12\dot\phi_0^2 + V \right) \, ,
\\
& {\dot H} = -\frac{\dot\phi_0^2}{2\mpl^2} \, ,
\\
\label{eq:bgphi}
& \ddot\phi_0^a + \Gamma^a_{bc}\dot\phi_0^b\dot\phi_0^c + 3H{\dot\phi_0}^a + {\cal G}^{ab}V_b = 0 \, ,
\end{align}
where $\dot\phi_0^2 \equiv \calG_{ab}\dot\phi_0^a\dot\phi_0^b$. Note that from the linear order action 
\begin{align}
S_1 = \int d^4x \bar{N} \bigg[ & \frac{a}{6} h \left( -\dot{P}+HP+\frac{P^2}{12 \mpl^2 a^2}+\frac{3{\cal G}^{ab}P_aP_b}{2a^4}-3a^2 V \right)
+ \frac{P}{3}\pi \left( \dot{a}+\frac{P^2}{6\mpl^2 a} \right)
\nonumber\\
& + \varphi^a \left( \dot{P}_a+\frac{1}{2a^3}{\cal G}^{cd}{}_{,a}P_cP_d+a^3V_a \right) + \pi_a \left( \dot{\phi}_0^a-\frac{{\cal G}^{ab}P_b}{a^3} \right)
\nonumber\\
& + \frac{n}{\bar{N}} \left( \frac{P^2}{12\mpl^2a} - \frac{\calG^{ab}P_aP_b}{2a^3} - a^3V \right) \bigg] \, ,
\end{align}
where $h \equiv h^i{}_i$, we can immediately read the same background equations, which we obtained by perturbing the zeroth order action, as the constraints for the perturbation variables.

\subsection{Poisson brackets}
\label{subsec:poisson}

Quantization of quantum fluctuations comes from the Poisson brackets. For gravity, the Poisson bracket is defined as
\begin{equation}
\left\{ A, B \right\} \equiv \int d^4x \left[ \frac{\delta A}{\delta \gamma_{ij}(x)}\frac{\delta B}{\delta\Pi^{ij}(x)}-\frac{\delta B}{\delta \gamma_{ij}(x)}\frac{\delta A}{\delta\Pi^{ij}(x)} \right] \, ,
\end{equation}
to give the equal time commutation relation between $\gamma_{ij}$ and its canonical momentum $\Pi^{ij}$:
\begin{equation}
\left\{ \gamma_{ij} (t, \mathbi{x}), \Pi^{kl} (t, \mathbi{y}) \right\} = \frac{1}{2} \left( \delta_i^k \delta_j^l +\delta_i^l \delta_j^k \right) \delta^{(3)}(\mathbi{x}-\mathbi{y}) \equiv \delta_{ij}^{kl}\delta^{(3)}(\mathbi{x}-\mathbi{y}) \, ,
\end{equation}
which is equivalent to 
\begin{equation}
\left\{ h_{ij} (t, \mathbi{x}), \pi^{kl} (t, \mathbi{y}) \right\} = -\frac{\delta_{ij}^{kl}}{a^3\mpl^2H} \delta^{(3)}(\mathbi{x}-\mathbi{y}) \, .
\end{equation}
For the matter fields, the Poisson bracket defined by
\begin{equation}
\left\{ A, B \right\} \equiv \int d^4x \left[ \frac{\delta A}{\delta \phi^a(x)}\frac{\delta B}{\delta\Pi_a(x)}-\frac{\delta B}{\delta \phi^a(x)}\frac{\delta A}{\delta\Pi_a(x)} \right] 
\end{equation}
gives
\begin{equation}
\left\{ \phi^a(t,\mathbi{x}),\Pi_b(t,\mathbi{y}) \right\} = \left\{ \varphi^a(t,\mathbi{x}),\pi_b(t,\mathbi{y}) \right\} = \delta^a_b \delta^{(3)}(\mathbi{x}-\mathbi{y}) \, .
\end{equation}
From these Poisson brackets, one can check the constraints ${\cal H}$ and ${\cal H}^i$ form a closed algebra~\cite{commutator} as follows.

Since the matter part of the action does not contain $\Pi^{ij}$, we can treat the gravity and matter parts separately. For the gravity part, we note that ${\cal H}_G^i$ satisfies
\begin{align}
\left\{ \gamma_{ij}(t, \mathbi{x}), \int d^3 x' \gamma_{kl}{\cal H}^k_G \xi^l (t, \mathbi{x}^\prime) \right\} & = \gamma_{ij,k}\xi^k+\gamma_{kj}\xi^k_{,i}+\gamma_{ik}\xi^k_{,j} \, ,
\\
\left\{ \Pi^{ij}(t, \mathbi{x}), \int d^3 x' \gamma_{kl}{\cal H}^k_G \xi^l (t, \mathbi{x}^\prime) \right\} & = (p^{ij} \xi^k)_{,k}-p^{kj}\xi^i_{,k}-p^{ik}\xi^j_{,k} \, ,
\end{align}
for an arbitrary infinitesimal parameter for spatial diffeomorphism $\xi^i(x)$ for $x^i \to x^i + \xi^i$. This implies that ${\cal H}_G^i$ are the generators of 3-dimensional diffeomorphism. Then, the Poisson brackets of two ${\cal H}_G^i$s at an equal time are just given by the structure constant of diffeomorphism. Thus, given two diffeomorphisms
\dis{
& x^i \to {x'}_A^i = x^i+ \xi_1^i \to {x''}_A^i={x'}_A^i+ \xi_2^i \, ,
\\
& x^i \to {x'}_B^i =x^i+ \xi_2^i \to {x''}_B^i={x'}_B^i+ \xi_1^i \, ,
}
Their commutator can be read off from
\dis{\gamma_{ij}&(x\to x'_A\to x''_A)-\gamma_{ij}(x\to x'_B\to x''_B)
\\
&=-\gamma_{ij,k} \left( \xi_1^l \xi^k_{2,l} - \xi_2^l \xi^k_{1,l} \right) - \gamma_{kj} \left( \xi_1^l \xi^k_{2,l} - \xi_2^l \xi^k_{1,l} \right)_{,i} - \gamma_{ik} \left( \xi_1^l \xi^k_{2,l} - \xi_2^l \xi^k_{1,l} \right)_{,j} \, ,}
such that
\begin{equation}
\left\{ \calH_{Gi}(t,\mathbi{x}), \calH_{Gj}(t, \mathbi{y}) \right\} = \calH_{Gi}(t,\mathbi{y})\partial_j^x \delta^{(3)}(\mathbi{x}-\mathbi{y}) - \calH_{Gj}(t,\mathbi{x})\partial_i^y\delta^{(3)}(\mathbi{x}-\mathbi{y}) \, .
\end{equation}
Since $\calH_G$ is a scalar with respect to diffeomorphism, we find
\begin{equation}
\left\{ \calH_{G}(t,\mathbi{x}), \calH_{Gi}(t,\mathbi{y}) \right\} = -\calH_G \partial_i \delta^{(3)}(\mathbi{x}-\mathbi{y}) \, .\
\end{equation}
For the commutator between two $\calH_G$s, using
\dis{\delta(\sqrt{\gamma}R)=\sqrt{\gamma}\gamma^{ij}\gamma^{kl} \left( \delta\gamma_{il;jk}-\delta \gamma_{ij;kl} \right) - \sqrt{\gamma} \left( R^{ij}-\frac12\gamma^{ij}R \right) \delta\gamma_{ij} \, ,}
we have 
\begin{equation}
\left\{ \calH_G(t,\mathbi{x}), \calH_G(t,\mathbi{y}) \right\} = \calH_G^i(t,\mathbi{y})\partial_i^x\delta^{(3)}(\mathbi{x}-\mathbi{y}) - \calH_G^i(t,\mathbi{x})\partial_i^y\delta^{(3)}(\mathbi{x}-\mathbi{y}) \, .
\end{equation}
For the matter part, similarly we obtain
\begin{equation}
\begin{split}
\left\{ \calH_{Mi}(t,\mathbi{x}), \calH_{Mj}(t, \mathbi{y}) \right\} & = \calH_{Mi}(t,\mathbi{y})\partial_j^x \delta^{(3)}(\mathbi{x}-\mathbi{y}) - \calH_{Mj}(t,\mathbi{x})\partial_i^y\delta^{(3)}(\mathbi{x}-\mathbi{y}) \, ,
\\
\left\{ \calH_{M}(t,\mathbi{x}), \calH_{Mi}(t,\mathbi{y}) \right\} & = -\calH_M(t,\mathbi{x}) \partial_i^x\delta^{(3)}(\mathbi{x}-\mathbi{y}) \, ,
\\
\left\{ \calH_M(t,\mathbi{x}), \calH_M(t,\mathbi{y}) \right\} & = \calH_M^i(t,\mathbi{y})\partial_i^x\delta^{(3)}(\mathbi{x}-\mathbi{y}) - \calH_M^i(t,\mathbi{x})\partial_i^y\delta^{(3)}(\mathbi{x}-\mathbi{y}) \, . 
\end{split}
\end{equation}
Thus, the total constraints $\calH = \calH_G + \calH_M$ and $\calH_i = \calH_{Gi}+\calH_{Mi}$ also form the closed algebra:
\begin{equation}
\begin{split}
\left\{ \calH_{i}(t,\mathbi{x}), \calH_{j}(t, \mathbi{y}) \right\}
& = \calH_{i}(t,\mathbi{y})\partial_j^x \delta^{(3)}(\mathbi{x}-\mathbi{y}) - \calH_{j}(t,\mathbi{x})\partial_i^y\delta^{(3)}(\mathbi{x}-\mathbi{y}) \, ,
\\
\left\{ \calH(t,\mathbi{x}), \calH_{i}(t,\mathbi{y}) \right\} & = -\calH \partial_i \delta(\mathbi{x}-\mathbi{y}) \, ,
\\
\left\{ \calH(t,\mathbi{x}), \calH(t,\mathbi{y}) \right\} & = \calH^i(t,\mathbi{y})\partial_i^x\delta^{(3)}(\mathbi{x}-\mathbi{y}) - \calH^i(t,\mathbi{x})\partial_i^y\delta^{(3)}(\mathbi{x}-\mathbi{y}) \, ,
\end{split}
\end{equation}
and they are interpreted as generators of diffeomorphisms in the time and space directions. Hereafter, for simplicity, we consider the case of flat field space, $\calG_{ab}=\delta_{ab}$. More general field space metric would introduce terms that contain the field space curvature, and replace the usual partial derivatives by covariant ones with respect to $\calG_{ab}$~\cite{Gong:2011uw}.

\section{Action for cosmological perturbations}
\label{Sec:quantization}
\setcounter{equation}{0}

\subsection{Gauge fixing}
\label{subsec:gaugefix}

Expanding the action around the classical solutions,  we obtain the action in the schematic form of
\begin{equation}
S = \int d^4x  \left( \pi_a \partial_t \varphi^a-2a^3 H \pi^{ij}\partial_t h_{ij}-{\cal H}+nC^0+N_i C^i \right) \equiv S_\text{free} + S_\text{int} \, ,
\label{Eq:Lagrangian0}
\end{equation}
with the free and interaction parts being given by
\dis{
S_\text{free} & = \int d^4x  \left( \pi_a \partial_t \varphi^a-2a^3 H \pi^{ij}\partial_t h_{ij} - \calH_\text{free} +n C^0_1+N_i C^i_1 \right) \, ,
\\
S_\text{int} & = \int d^4 x  \left( -\calH_\text{int} +n C^0_{\geq 2}+ N_i C^i_{\geq 2} \right) \, ,
\label{Eq:Lagrangian}
}
where $\calH_\text{free}$ and $\calH_\text{int}$ denote respectively the Hamiltonian quadratic in perturbations and cubic and beyond, and $C^\mu_1$ and $C^\mu_{\geq2}$ denote the constraints linear in perturbations and quadratic and beyond, respectively. We present these terms in detail in Appendix~\ref{app:lagrangian}. Since gravity has four constraints, we need four gauge fixing conditions $\psi_\mu$ ($\mu=0,1,2,3$) satisfying
\begin{equation}
{\rm det}\left( \left\{C^\mu,\psi_\nu\right\} \right)\ne 0 \, ,
\label{Eq:FPcondition}
\end{equation}
where $C^\mu=(C^0, C^i)$ are the constraints, or the generators of diffeomorphisms in  \eqref{Eq:Lagrangian0}. More explicitly, the constraints $C^\mu$ have the following Poisson brackets with the fluctuations $h_{ij}$ and $\varphi^a$:
\begin{align}
\left\{ h_{ij}(t, \mathbi{x}), C^0(t, \mathbi{y}) \right\} & = -2H{\tilde\gamma}^{-1/2} \Big\{ \delta_{ij} + \left( -h_{ij}+\delta_{ij}h-2\pi^{ij}+\delta_{ij}\pi \right)
\nonumber\\
& \qquad\qquad\qquad + \left[ (h+\pi)h_{ij}-2h_{ik}h_{jk}-2(h_{il}\pi^{jl}+h_{jl}\pi^{il})+\pi^{kl}h_{kl}\delta_{ij} \right]
\nonumber\\
& \qquad\qquad\qquad + (-2 h_{il}\pi^{lk}h_{jk}+h_{kl}\pi^{kl}h_{ij}) \Big\} \delta^{(3)}(\mathbi{x}-\mathbi{y}) \, ,
\\
\left\{ h_{ij}(t, \mathbi{x}), C^k(t, \mathbi{y}) \right\} & = -\frac{2}{a^2} \left( \delta_{k(i}\partial^x_{j)}-\Gamma^k_{ij} \right) \delta^{(3)}(\mathbi{x}-\mathbi{y}) \, ,
\\
\left\{ \varphi^a(t, \mathbi{x}), C^0(t, \mathbi{y}) \right\} & = -\frac{\tilde{\gamma}^{-1/2}}{a^3}(P_a+\pi_a) \delta^{(3)}(\mathbi{x}-\mathbi{y}) \, ,
\\
\left\{ \varphi^a(t, \mathbi{x}), C^i(t, \mathbi{y}) \right\} & = -\frac{1}{a^3}\tilde{\gamma}^{ik}\partial_k\varphi^a \delta^{(3)}(\mathbi{x}-\mathbi{y}) \, ,
\end{align}
where $\tilde\gamma_{ij} = \gamma_{ij}/a^2$. In principle, we can think of four possibilities:
\begin{itemize}
 \item[1.] Both the $\psi^0$ and the $\psi^i$ conditions come from $h_{ij}$.
 \item[2.] Both the $\psi^0$ and the $\psi^i$ conditions come from $\varphi^a$.
 \item[3.] The $\psi^0$ condition comes from $h_{ij}$, whereas the $\psi^i$ condition comes from $\varphi^a$.
 \item[4.] The $\psi^0$ condition comes from $\varphi^a$, whereas the $\psi^i$ condition comes from $h_{ij}$.  
\end{itemize}
However, the 3-vector condition made up of $\varphi^a$, say, $\psi_i=\psi_i(\varphi^a,  \partial_k \varphi^a)$ satisfies
\begin{align}
\left\{ \psi_j(\varphi)(t, \mathbi{x}), C^i(t, \mathbi{y}) \right\} & = -\frac{\tilde{\gamma}^{ik}}{a^2}  \partial_k \varphi^a (t, \mathbi{y}) \left[ \frac{\partial\psi_j}{\partial \varphi^a}(t, \mathbi{x}) + \frac{\partial \psi_j}{\partial (\partial_l\varphi^a)}(t, \mathbi{x})\partial^x_l \right] \delta^{(3)}(\mathbi{x}-\mathbi{y}) 
\nonumber\\
&= -\frac{\tilde\gamma^{ik}}{a^2} D_k\psi_j \delta^{(3)}(\mathbi{x}-\mathbi{y}) \, .
\end{align}
Thus it vanishes under $\psi_i=0$, violating \eqref{Eq:FPcondition}: the vector constraints $C^i$ cannot be satisfied with $\varphi^a$. Hence, plausible gauge fixing conditions should be chosen between the possibilities 1 and 4.

Let us consider the first possibility in which the gauge fixing is imposed entirely from the metric fluctuations. One simple example is to take
\begin{equation}
\label{eq:flatgauge}
\psi_0=h \quad \text{and} \quad \psi_i = \partial^j \left( h_{ij}-\frac{\delta_{ij}}{3}h \right) \, .
\end{equation}
As we will see, in terms of the metric decomposition in \eqref{Eq:metricdec}, the gauge fixing by $\psi_\mu=0$ is equivalent to setting the scalar components of $h_{ij}$ zero. Thus the scalar degrees of freedom are entirely given to $\varphi^a$, which is the so-called ``flat'' gauge condition usually taken in multi-field inflation. On the other hand, as an example of the fourth possibility,  we can take the gauge fixing conditions as
\begin{equation}
\label{eq:comgauge}
\psi_0=\delta_{ab}\dot{\phi}_0^a \varphi^b \quad \text{and} \quad \psi_i= \partial^j \left( h_{ij}-\frac{\delta_{ij}}{3}h \right) \, .
\end{equation}
Here, the condition $\psi_0=0$ is equivalent to setting $T^0{}_i=0$, so it is known as the ``comoving'' gauge condition. A more convenient way of implementing $\psi_0=0$ will be discussed in the following sections.

For completeness, we list the Poisson brackets of the gauge fixing conditions \eqref{eq:flatgauge} and \eqref{eq:comgauge} before imposing $\psi_\mu=0$ as follows:
\begin{align}
\left\{ h(t, \mathbi{x}), C_0(t, \mathbi{y}) \right\} & = -2H\tilde{\gamma}^{-1/2} \Big[ 3+(2h+\pi) + \left( h^2+\pi h-2h_{ij}h_{ij}-\pi^{ij}h_{ij} \right)
\nonumber\\
& \qquad\qquad\qquad + \left( -2h_{ik}\pi^{kl}h_{li}+h \pi^{ij}h_{ij} \right) \Big] \delta^{(3)}(\mathbi{x}-\mathbi{y}) \, , 
\label{Eq:hscalar1}
\\
\left\{ h(t, \mathbi{x}), C_i(t, \mathbi{y}) \right\} & = -\frac{2}{a^2} \left[ \partial_i^x-\tilde{\gamma}^{ij} 
\left(\partial^kh_{jk}-\frac12 h_{,j}\right) 
\right] \delta^{(3)}(\mathbi{x}-\mathbi{y}) \, ,
\label{Eq:hscalar2}
\\
\left\{ \delta_{ab} \dot{\phi}_0^a \varphi^b(t, \mathbi{x}), C_0(t, \mathbi{y}) \right\} & = -\frac{\tilde{\gamma}^{-1/2}}{a^3} \dot{\phi}_0^a(P_a+\pi_a)(t, \mathbi{y}) \delta^{(3)}(\mathbi{x}-\mathbi{y}) \, ,
\label{Eq:phiscalar1}
\\
\left\{ \delta_{ab}  \dot{\phi}_0^a \varphi^b(t, \mathbi{x}), C_i(t, \mathbi{y}) \right\} & = -\frac{1}{a^2}\tilde{\gamma}^{ik}\delta_{ab}  \dot{\phi}_0^a\partial_k \varphi^b(t, \mathbi{y}) \delta^{(3)}(\mathbi{x}-\mathbi{y}) \, ,
\label{Eq:phiscalar2}
\\
\left\{ \partial_j \left( h_{ij}-\frac{\delta_{ij}}{3}h \right)(t, \mathbi{x}), C_0(t, \mathbi{y}) \right\} & = -2H\tilde{\gamma}^{-1/2} \Big[ - \left( h_{ij}-\frac{\delta_{ij}}{3}h \right) - 2 \left( \pi^{ij}-\frac{\delta_{ij}}{3}\pi \right) 
\nonumber\\
& \qquad\qquad\qquad  + (h+\pi)\left( h_{ij}-\frac{\delta_{ij}}{3}h \right) - 2 \left( h_{ik}h_{jk}-\frac{\delta_{ij}}{3}h_{kl}h_{kl} \right) 
\nonumber \\
& \qquad\qquad\qquad - 2\left( h_{ik}\pi^{jk}+h_{jk}\pi^{ik}-\frac23\delta_{ij}\pi^{kl}h_{kl} \right) 
\nonumber\\
& \qquad\qquad\qquad  -2 \left( h_{il}\pi^{lk}h_{kj}-\frac{\delta_{ij}}{3}h_{lm}\pi^{lk}h_{km} \right) 
\nonumber\\
& \qquad\qquad\qquad 
\left. + \pi^{kl}h_{kl} \left( h_{ij}-\frac{\delta_{ij}}{3}h \right) \right]  (t, \mathbi{y})\partial_j^x \delta^{(3)}(\mathbi{x}-\mathbi{y}) \, ,
\label{Eq:hvector1}
\\
\left\{ \partial_j \left( h_{ij}-\frac{\delta_{ij}}{3}h \right)(t, \mathbi{x}), C_j(t, \mathbi{y}) \right\} & = -\frac{1}{a^2} \Big[ \delta_{ij}\nabla^2_x+\frac13 \partial_i^x \partial_j^x - \tilde{\gamma}^{jk} \left( 2\partial_{(i}h_{l)k}-\partial_k h_{il} \right)(t,\mathbi{y})\partial_l^x
\nonumber\\
& \qquad\quad + \frac13\tilde{\gamma}^{jk} \left( 2\partial_lh_{lk}-\partial_k h \right)(t, \mathbi{y})\partial_i^x \Big] \delta^{(3)}(\mathbi{x}-\mathbi{y}) \, .
\label{Eq:hvector2}
\end{align}

\subsection{Quadratic action}
\label{subsec:quadratic}

Now we are ready to study the perturbed action. We begin with the quadratic action, for which we decompose the metric perturbation $h_{ij}$ using the scalar, vector and tensor components in the standard manner as
\begin{equation}
\label{Eq:metricdec}
h_{ij} = 2H_L\delta_{ij} + 2\left( \partial_i\partial_j - \frac{\delta_{ij}}{3}\nabla^2 \right)H_T + \partial_{(i}h_{j)}^T + h_{ij}^{TT} \, ,
\end{equation}
where the vector $h_i^T$ is transverse, and the tensor $h_{ij}^{TT}$ is transverse and traceless. Defining the multi-field version of the gauge-invariant Mukhanov-Sasaki variable~\cite{Anderegg:1994xq,MSvariable} as
\begin{equation}
\tilde{\varphi}^a \equiv \varphi^a - \frac{\dot{\phi}_0^a}{H} \left( H_L-\frac{\nabla^2}{3} H_T \right) \, ,
\label{Eq:multiredef}
\end{equation}
the quadratic action is greatly simplified to give
\begin{align}
S_2 & = \int d^3x \bar{N}dt \frac{a^3}{2} \left\{ \delta_{ab}\dot{\tilde{\varphi}}^a\dot{\tilde{\varphi}}^b - \frac{\delta_{ab}}{a^2}\partial^i \tilde{\varphi}^a\partial_i \tilde{\varphi}^b - M_{ab}^2 \tilde{\varphi}^a\tilde{\varphi}^b + \frac{\mpl^2}{4} \left[ \left( \dot{h}_{ij}^{TT} \right)^2 - \frac{1}{a^2} \partial^k{h}_{ij}^{TT} \partial_k{h}_{ij}^{TT} \right] \right\}
\nonumber\\
& \quad + \text{(auxiliary field terms)} \, ,
\label{Eq:quadratic} 
\end{align}
where
\begin{equation}
M_{ab}^2 = V_{ab}+\frac{V}{\mpl^4H^2}\delta_{ac}\delta_{bd}\dot\phi_0^c\dot\phi_0^d + \frac{1}{\mpl^2 H}(\delta_{ac}V_b+\delta_{bc}V_a)\dot\phi_0^c \, .
\end{equation} 
Note that the auxiliary field terms contain those quadratic in momentum and the unphysical fields, which are separated from the quadratic action of dynamical fields after appropriate redefinitions. However, these redefined auxiliary fields appear in the higher order action through, for example, ghost terms as can be seen in Appendix~\ref{app:ghost}. Even though we have to integrate them out in the path integral, as asserted in the beginning of Appendix~\ref{app:ghost}, it is quite convenient to leave them as auxiliary fields. We list them in Appendix~\ref{App:quadmom}.

The pure tensor action in \eqref{Eq:quadratic} is precisely the same as that in single-field inflation, so from now on we concentrate on the scalar action. If we take the flat gauge \eqref{eq:flatgauge}, which amounts to $H_L = H_T = 0$ in \eqref{Eq:metricdec}, from \eqref{Eq:multiredef} we have simply $\tilde\varphi^a = \varphi^a$. Then, \eqref{Eq:quadratic} is reduced to the well-known quadratic scalar action~\cite{Anderegg:1994xq,Gong:2011uw,Langlois:2008mn,Gao:2012uq,Gao:2015aba}. Taking the flat gauge is conventional for multi-field inflation: there are in total $N$ physical degrees of freedom after eliminating the unphysical ones, thus it is very natural to assign them to the $N$ field fluctuations. Meanwhile, it is more difficult to implement the comoving gauge in which the notion of the comoving curvature perturbation [see \eqref{eq:R}] becomes manifest, because naively all $N$ field fluctuations contribute to fix the temporal gauge condition $C^0$ as can be read from the condition for $\psi_0$ in \eqref{eq:comgauge}. Thus, \eqref{Eq:multiredef} is, as it is, not appropriate to apply the comoving gauge. Instead, it is more meaningful to decompose $\varphi^a$ into the directions along and orthogonal to time evolution~\cite{Achucarro:2012sm} as 
\begin{equation}
\label{eq:decomposition2}
\varphi^a(t,\mathbi{x}) = \varphi^a_\bot(t,\mathbi{x}) + \dot\phi^a_0(t)\tilde\pi(t,\mathbi{x}) \, ,
\end{equation}
with the orthogonality condition $\delta_{ab}\dot\phi^a_0\varphi_\bot^b = 0$. Then the temporal gauge condition of \eqref{eq:comgauge} can be  rewritten as
\begin{equation}
\psi_0 = \dot\phi_0^2\tilde\pi \, .
\end{equation}
Thus, the comoving gauge is simply imposed upon $\tilde\pi=0$. Note that the linear gauge transformation $\varphi^a \to \varphi^a - \dot\phi^a_0\xi^0$ tells us
\begin{equation}
\tilde\pi \to \tilde\pi - \xi^0 \quad \text{and} \quad \varphi^a_\bot \to \varphi^a_\bot \, .
\end{equation}
This is the restatement that $\tilde\pi$ is the fluctuation in the direction of the time translation, and is thus interpreted as the Goldstone mode resulting from the spontaneous breaking of the time translation invariance~\cite{Cheung:2007st}. Meanwhile, the orthogonal fluctuations, which are usually called ``isocurvature'' modes, are gauge invariant~\cite{Gordon:2000hv}. Then, \eqref{Eq:multiredef} is now written as
\begin{equation}
\tilde\varphi^a(t,\mathbi{x}) = \varphi^a_\bot(t,\mathbi{x}) - \frac{\dot\phi^a_0}{H} \left( H_L - \frac{\nabla^2}{3}H_T - H\tilde\pi \right)(t,\mathbi{x}) \equiv \varphi^a_\bot(t,\mathbi{x}) - \frac{\dot\phi^a_0}{H} \pi(t,\mathbi{x}) \, .
\end{equation}
Note that due to the orthogonality condition, $N-1$ out of $N$ $\varphi_\bot^a$'s are independent in the comoving gauge: the remaining single degree of freedom is $\pi$, which is also gauge invariant. With this decomposition, the scalar quadratic action is rewritten in terms of gauge invariant variables $\pi$ and $\varphi^a_\bot$ as
\begin{align}
\label{eq:S2scalar}
S_2^{(s)} = \int d^3x \bar{N}dt \frac{a^3}{2} \left[ \delta_{ab}\dot\varphi_\bot^a\dot\varphi_\bot^b - \frac{\delta_{ab}}{a^2}\partial^i\varphi_\bot^a\partial_i\varphi_\bot^b - V_{ab}\varphi_\bot^a\varphi_\bot^b + 2\epsilon\mpl^2 \left( \dot\pi^2 - \frac{1}{a^2}\partial^i\pi\partial_i\pi \right) - \frac{4}{H}V_a\varphi^a_\bot\dot\pi \right] \, ,
\end{align}
where for the last term we have used the relation $\delta_{ab}\dot\phi^a_0\dot\varphi^b_\bot = V_a\varphi_\bot^a$.

Note that if we do not have any light mode other than $\pi$, we can integrate out the heavy isocurvature modes $\varphi_\bot^a$ by performing the path integral over them as
\begin{equation}
\int {\cal D}\varphi_\bot^a \exp \left( -\frac12 {\varphi}^a_\bot D^{\bot}_{ab}\varphi_\bot^b-J_a\varphi_\bot^a \right) = \exp \left[ \frac12 J_a \left(D^{\bot -1}\right)^{ab}J_b \right]
\end{equation}
for certain operators $D^{\bot}_{ab}$ and $J_a$.
We then obtain the effective single field action for $\pi$ as
\begin{equation}
S_{\rm eff} = \int d^3x \bar{N}dt a^3\epsilon\mpl^2 \left[ \left\{ 1 + \frac{2}{\epsilon\mpl^2H^2} V_a \left[ \left( M_{\rm eff}^2 \right)^{-1} \right]^{ab}V_b \right\} \dot\pi^2 - \frac{1}{a^2} (\nabla\pi)^2 \right] + \cdots \, ,
\end{equation}
where dots denotes non-local terms. The effective mass matrix of $\varphi_\bot^a$, $\left( M_{\rm eff}^2 \right)_{ab}$, is given by
\begin{equation}
\left( M_{\rm eff}^2 \right)_{ab} = \left( R^{-1}V''R \right)_{ab} \, , \quad (a,b=2,\cdots, N)
\end{equation}
with $R$ being a unitary matrix rotating the unit vector $(1,0,\cdots, 0)$ to $1/(\mpl H \sqrt{2\epsilon}) \left( \dot{\phi}^1_0,\cdots \dot{\phi}^N_0 \right)$~\cite{Gao:2012uq,Gao:2015aba,Burgess:2012dz}. The coefficient of $\dot\pi^2$ is interpreted as the speed of sound $c_s^{-2}$~\cite{Achucarro:2012sm,Achucarro:2010da}, which is reduced by the interaction with $\varphi^a_\bot$, as the kinetic energy of $\pi$ is extracted to excite $\varphi^a_\bot$.

In perfect de Sitter spacetime where $H$ is constant, time translation is not broken up to dilatation. To state this mathematically, de Sitter spacetime has a conformal Killing vector for time translation, which belongs to the $SO(4,2)$ conformal isometry, even though it does not belong to the $SO(4,1)$ isometry. On the other hand, in order to exit the inflationary phase, $H$ should be time-dependent such that the background distinguishes different time slices through time evolution. In this sense, $\dot{H}$, or equivalently the slow-roll parameter $\epsilon$, is the order parameter for spontaneous breaking of time translation invariance. Since $\epsilon \propto \dot\phi_0^2$, we can say that time evolution of classical solution $\phi_0(t)$ distinguishes different time slices as well. In this regard, the factor $2\epsilon\mpl^2$ in the quadratic action is an analogy of the pion decay constant in chiral perturbation theory, which parametrizes the spontaneous breaking of $SU(3) \times SU(3)$ chiral symmetry.

The mass of $\pi$ is obtained by investigating the redefined field $\sqrt{2\epsilon}\mpl\pi$ in the canonical basis: after integration by parts, its mass squared is given by
\dis{ -m_\pi^2=\frac{3H}{\sqrt{\epsilon}}\frac{d\sqrt{\epsilon}}{dt}+\frac{1}{\sqrt{\epsilon}}\frac{d^2\sqrt{\epsilon}}{dt^2}=
   3H\frac{\dot{\epsilon}}{2\epsilon}+\frac{\ddot{\epsilon}}{2\epsilon}-\frac{\dot{\epsilon}^2}{4\epsilon^2} \, .}
Introducing another slow-roll parameter $\delta$ as
\dis{\delta \equiv -\frac{\delta_{ab}\dot{\phi}_0^a \ddot{\phi}_0^b}{H\dot{\phi}_0^2} \, ,}
in the slow-roll approximation $1\gg \epsilon \gg \dot{\epsilon}/H \gg \ddot{\epsilon}/H^2 \cdots$, we have $m_\pi^2 \approx -3H\dot{\epsilon}/(2\epsilon)=-3(\epsilon-\delta)H^2$. For energy much higher than $\sqrt{\epsilon} H$, it is a good approximation to set $\pi$ as a massless field, which is what we expect from the Goldstone equivalence theorem.

Before closing this section, we recall the notion of the curvature perturbation in the comoving gauge in a form useful for higher order correlation function discussed later. In the comoving gauge $\tilde{\pi}=0$, the natural perturbation variable of the metric $\gamma_{ij}$ is the ``local expansion factor'' $\calR$ defined by~\cite{Lyth:2004gb}
\begin{equation}
\label{eq:R}
{\rm det}(\gamma_{ij})|_{\tilde{\pi}=0} = \left[ a^2(t) e^{2\calR(t,\mathbi{x})} \right]^3 \, ,
\end{equation}
which also denotes the curvature perturbation in the otherwise homogeneous spatial hypersurfaces. The transformation from $\pi$ to $\calR$ is given by~\cite{Maldacena:2002vr,Noh:2004bc}
\begin{align}
\label{eq:pi->R}
\calR & = \pi + \left( \epsilon-\frac{\delta}{2} \right)\pi^2 + \frac{1}{H}\pi \dot{\pi} - \frac{1}{4a^2H^2} \left[ \partial^i\pi\partial_i\pi - \frac{\partial^i\partial^j}{\nabla^2} \left( \partial_i\pi\partial_j\pi \right) \right]
\nonumber\\
& \quad + \frac{\epsilon}{H} \left[ \partial^i\pi\frac{\partial_i}{\nabla^2}\dot{\pi} - \frac{\partial^i\partial^j}{\nabla^2} \left( \partial_i\pi\frac{\partial_j}{\nabla^2} \dot{\pi} \right) \right] - \frac{1}{4H}\dot{h}_{ij}^{TT}\partial^i\partial^j \pi + \cdots \, .
\end{align}
In terms of $\calR$, the quadratic action remains the same with $\pi$ replaced with $\calR$. But the transformation \eqref{eq:pi->R} does give rise to additional contributions to the higher order action in terms of $\pi$ as we will see in Section~\ref{Sec:higherorder}.

\section{Two-point correlation functions}
\label{Sec:2point}
\setcounter{equation}{0}

\subsection{In-in formalism}

In the study of cosmological perturbations, the in-in formalism, or the Schwinger-Keldysh formalism~\cite{Maldacena:2002vr,in-in} concerns with expectation values of operators sandwiched between the vacuum at a specific time $| \Omega, t_i \rangle$, usually taken at $t_i \to -\infty$. It is different from the in-out formalism for scattering processes in particle physics, in which we investigate the evolution of operators from an in-state ($t_i \to -\infty$) to an out-state ($t_f \to +\infty$).  In the in-in formalism, the generating functional is given by
\dis{Z[J_{+}, J_{-}]&=\sum_\alpha \langle \Omega, t_i | \alpha, t_f \rangle_{J_{-}} \langle \alpha, t_f | \Omega, t_i \rangle_{J_{+}}
\\
&=\int {\cal D}\Phi_+{\cal D}\Phi_- e^{iS[\Phi_+]-iS[\Phi_-]+i J_+\Phi_+-iJ_-\Phi_-} \, ,
}
where $\Phi$ denotes fields in the action collectively, and $t_i$ ($t_f$) is usually taken as $-\infty$ ($+\infty$). From this, we have the Green's functions
 \dis{G^{+\cdots + -\cdots -}(x_1,\cdots, x_n, y_1, \cdots, y_m)&=\langle \Omega, t_i | \bar{T}[\Phi(y_1) \cdots  \Phi(y_m)]T[\Phi(x_1) \cdots \Phi(x_n)]|\Omega, t_i\rangle
 \\
&=\prod_{j=1}^m\frac{\delta}{-i\delta J_-(y_j)}\prod_{i=1}^n \frac{\delta}{i\delta J_+(x_i)}Z[J_+, J_-]\Big|_{J_+=J_-=0} \, ,
}
where the time-ordering and anti time-ordering operators give
\dis{&T[\Phi(x)\Phi(x')]=\theta(x^0-x^{\prime 0})\Phi(x)\Phi(x')+\theta(x^{\prime 0}-x^0)\Phi(x')\Phi(x) \, ,
\\
&\bar{T}[\Phi(x)\Phi(x')]=\theta(x^{\prime 0}-x^0)\Phi(x)\Phi(x')+\theta(x^0-x^{\prime 0})\Phi(x')\Phi(x) \, .}

As the first step to obtain the Feynman rules, we separate the action into the quadratic piece and  the interaction terms~\cite{Prokopec:2010be},
\dis{S[\Phi]=\int d^4x \frac12 \Phi D \Phi+S_\text{int}[\Phi] \, ,}
such that the generating functional is written as the derivative of the generating functional:
\dis{Z[J_+, J_-] & = e^{iS_I \left[ \frac{\delta}{i\delta J_+} \right]-iS_I\left[ \frac{\delta}{-i\delta J_-} \right]} \int {\cal D}\Phi_+{\cal D}\Phi_- \exp \left\{ \int d^4x \left[ \frac{i}{2}
\left( \begin{array}{cc}
\Phi_+ & \Phi_-
\\
\end{array}\right)
\left( \begin{array}{cc}
D& 0\\
0 & -D
\end{array}\right)
\left( \begin{array}{c}
\Phi_+
\\
\Phi_-
\end{array}\right)
\right.\right.
\\
& \hspace{17em} \left.\left. +
i\left( \begin{array}{cc}
\Phi_+ & \Phi_-
\\
\end{array}\right)
\left( \begin{array}{c}
J_+\\
-J_-
\end{array}\right)
\right] \right\}
\\
& = e^{iS_I\left[ \frac{\delta}{i\delta J_+} \right]-iS_I\left[ \frac{\delta}{-i\delta J_-} \right]} \int {\cal D}\Phi_+{\cal D}\Phi_- \exp \left[ \int d^4x d^4x'\frac{-1}{2}
\left( \begin{array}{cc}
J_+& -J_-\\
\end{array}\right)(x)
\right.
\\
& \hspace{17em} \left.
\times\left( \begin{array}{cc}
i\triangle_{++}& i\triangle_{+-}
\\
i\triangle_{-+} & i\triangle_{--}
\end{array}\right)(x,x')
\left( \begin{array}{c}
J_+\\
-J_-
\end{array}\right)(x') \right] \, ,
}
where the propagators satisfy
\dis{
\left( \begin{array}{cc}
D& 0
\\
0 & -D
\end{array}\right)(x)
\left( \begin{array}{cc}
i\triangle_{++}& i\triangle_{+-}
\\
i\triangle_{-+} & i\triangle_{--}
\end{array}\right)(x,x')
=i\delta^{(4)}(x-x') \, .
}

The four propagators $\triangle_{\pm\pm}$ are expressed in terms of the Wightman functions, 
\dis{&i\triangle^{>}(x,x')=\langle \Omega | \Phi(x) \Phi(x') |\Omega \rangle \, ,
\\
&i\triangle^{<}(x,x')=\langle \Omega | \Phi(x') \Phi(x)  |\Omega \rangle \, ,}
both of which are solutions to the operator equation
\dis{D(x)i\triangle^{\gtrless}(x,x')=0 \, .}
From $\partial_{x^0} \theta(x^0-x^{\prime 0})=\delta(x^0-x^{\prime 0})$ and $\delta(x^0-x^{\prime 0})[\Phi(x), \Phi(x')]=0$, we find that
\begin{align}
i\triangle_{++}(x, x') & = \langle \Omega | T[\Phi(x)\Phi(x')] | \Omega \rangle
\nonumber\\
& = \theta(x^0-x^{\prime 0})i\triangle^>(x, x')+\theta(x^{\prime 0}-x^0)i\triangle^<(x,x') \, ,
\\
i\triangle_{--}(x, x') & = \langle \Omega | \bar{T}[\Phi(x)\Phi(x')] | \Omega \rangle 
\nonumber\\
& = \theta(x^{\prime 0}-x^0)i\triangle^>(x, x')+\theta(x^0-x^{\prime 0})i\triangle^<(x,x') \, .
\end{align}
On the other hand, the boundary conditions at $t'$ (typically taken at $+\infty$), $J_+=J_-$ and $\langle \Omega | \Phi_+ | \Omega \rangle =\langle \Omega | \Phi^- | \Omega \rangle$ are imposed, such that
\dis{i\triangle_{+-}(x, x') & = i\triangle ^<(x, x') \, ,
\\
i\triangle_{-+}(x, x') & = i\triangle ^>(x, x') \, , }
satisfying  $i\Delta_{+-}=-i\Delta_{-+}^*$.

\subsection{Scalar and tensor propagators}

From the quadratic action \eqref{Eq:quadratic}, we can write down the differential equations that the propagators satisfy:
\dis{
\left[ \delta_{ab} \left( -\partial_t \frac{a^3}{\bar{N}}\partial_t + \bar{N}a \nabla^2 \right) - 2{\bar N}a^3 \left( 2V_{ab}+\frac{V}{2H^2}\delta_{ad}\delta_{be}\dot{\phi}^d \dot{\phi}^e+\frac{2}{H}\delta_{(a|d}V_{|b)} \dot{\phi}^d \right) \right] i\triangle^{bc}_{\pm \pm} & =\pm i \delta_a^c \delta^{(4)}(x-x') \, ,
\\
\left[ \delta_{ab} \left(-\partial_t \frac{a^3}{\bar{N}}\partial_t + \bar{N}a \nabla^2 \right) - 2{\bar N}a^3 \left( 2V_{ab}+\frac{V}{2H^2}\delta_{ad}\delta_{be}\dot{\phi}^d \dot{\phi}^e+\frac{2}{H}\delta_{(a|d}V_{|b)} \dot{\phi}^d \right) \right] i\triangle^{bc}_{\pm \mp} & = 0 
\label{Eq:scalarprop}
}
for scalars, and
\dis{
\left( -\partial_t \frac{a^3}{\bar{N}}\partial_t + \bar{N}a \nabla^2 \right)  i\triangle^{ijkl}_{\pm \pm} & = \pm i (P_{ik}P_{jl}+P_{il}P_{jk}-P_{ij}P_{kl})\delta^{(4)}(x-x') \, ,
\\
\left( -\partial_t \frac{a^3}{\bar{N}}\partial_t + \bar{N}a \nabla^2 \right)  i\triangle^{ijkl}_{\pm \mp} & = 0
\label{Eq:multiGreen}
}
for graviton, where $P_{ij}=\delta_{ij}-\partial_i \partial_j/\nabla^2$ is the projector in the transverse direction. Note that the scalar propagators have indices for different field contents [see \eqref{Eq:scalarprop}]. However, the leading action for the scalar perturbations is Gaussian, i.e.  there is no mixing between different scalars at leading order. Their mixings do contribute as corrections to the power spectrum which we will discuss shortly in Section~\ref{subsec:powerspec}. For this reason, let us first consider the free part of the quadratic action.

The free part of the quadratic action for the curvature perturbation $\calR$ and the graviton $h_{ij}^{TT}$ is schematically written in the form of
\begin{equation}
\label{eq:quadratic-free}
S_\text{free} = \int d^3x \bar{N}dt \frac{a^3}{2} s^2(t) \left[ \dot\Psi^2 - \frac{1}{a^2}\partial^i\Psi\partial_i\Psi - m^2(t)\Psi^2 \right] \, .
\end{equation}
In the case of $\calR$, $s^2=2\mpl^2\epsilon$ and $m^2=0$. For the graviton, the polarization factor $\sum_{\lambda=+,\times} \varepsilon^\lambda_{ij} \varepsilon^\lambda_{kl}$ is attached to the propagators in addition to $s^2=\mpl^2/4$ and $m^2=0$ for each polarization. Note that for the orthogonal components $\varphi_\bot^a$ and the field fluctuations $\varphi^a$ in the flat gauge, we can always extract the same free action as \eqref{eq:quadratic-free} for them by e.g. performing an appropriate rotation in the field space to make $V_{ab}$ diagonal~\cite{Gao:2012uq,Gao:2015aba,Burgess:2012dz} despite that in general we have non-trivial mixing between them via $M_{ab}^2$ (for $\varphi^a$) or $V_{ab}$ (for $\varphi_\bot^a$). Now suppose $\epsilon$ is a constant, which is a good approximation for $\calR$ as $\epsilon \gg \dot{\epsilon}/H$ in the slow-roll inflation. The resulting equation in terms of the conformal time $\tau$ from \eqref{eq:quadratic-free} for $\Psi$ is, with $a = -1/(H\tau)$,
\begin{equation}
\frac{d^2\Psi_k}{d\tau^2} - \frac{2}{\tau}\frac{d\Psi_k}{d\tau} + \left( k^2+m^2 \right)\Psi_k = 0 \, .
\end{equation}
In order to obtain $\Delta_{\pm\mp}$ for $\Psi$, we just replace $\Psi_k$ by the propagators. Solving this equation, we obtain the well-known Hankel function solution:
\begin{equation}
\label{Eq:massivesol}
\Psi_k = -\frac{i}{s} e^{i(\nu+1/2)\pi/2} \frac{\sqrt{\pi}}{2}H(-\tau)^{3/2} H_\nu^{(1)}(-k\tau) \quad \text{with} \quad \nu \equiv \sqrt{\frac94-\frac{m^2}{H^2}} \, .
\end{equation}
The corresponding dimensionless power spectrum is defined by
\begin{equation}
\calP_\Psi(k) \equiv \frac{k^3}{2\pi^2} |\Psi_k|^2 = \lim_{-k\tau\to0} \frac{k^3}{8\pi s^2} H^2 (-\tau)^3 \left| H_\nu^{(1)}(-k\tau) \right|^2 \, .
\end{equation}
Especially, in the massless limit $m \ll H$, the index of the Hankel function is just 3/2, and we find the power spectrum for a free massless field:
\begin{equation}
\calP_\Psi = \left( \frac{H}{2\pi s} \right)^2 \, .
\end{equation}

\subsection{Corrections to the power spectrum}
\label{subsec:powerspec}

Having found the free power spectrum of $\Psi$ in the previous section, now we consider the corrections due to the interaction terms between different fields, which we collectively denote by $\Phi$. Schematically we write the interaction terms as
\begin{equation}
S_\text{int} = \int d^3x \bar{N}dt \, a^3c(t) \calO^{(\Phi)} \Phi \calO^{(\Psi)} \Psi \, ,
\end{equation}
where $c(t)$ is the time-dependent coupling between $\Psi$ and $\Phi$, and $\calO^{(X)}$ is the possible derivative operator for the field $X$. Treating the interaction terms as perturbative interaction Hamiltonian, we can write the generating functional as
\begin{align}
Z[J_+, J_-] & = \int \calD\Psi_ + \calD\Psi_- e^{iS_{\rm free}[\Psi_+] - iS_{\rm free}[\Psi_-] + iJ_+\Psi_+ - iJ_-\Psi_-} e^{-i\int dt H_\text{int}^+ + i\int dt H_\text{int}^-}
\nonumber\\
& = \int \calD\Psi_+ \calD\Psi_- e^{iS_{\rm free}[\Psi_+] - iS_{\rm free}[\Psi_-] + iJ_+\Psi_+ - iJ_-\Psi_-}
\nonumber\\
& \qquad\quad \times \left[ 1 - i \left( \int dt H_\text{int}^+ - \int dt H_\text{int}^- \right) - \frac12 \left( \int dt H_\text{int}^+ - \int dt H_\text{int}^- \right)^2 + \cdots \right] \, .
\end{align}
From this, the two-point correlation function of $\Psi$ is given by
\begin{equation}
\left\langle \Omega \left| \Psi(t,\mathbi{x})\Psi(t,\mathbi{y}) \right| \Omega \right\rangle = \left. \frac{\delta}{i\delta J_+(t,\mathbi{x})} \frac{\delta}{i\delta J_+(t,\mathbi{x})} Z[J_+,J_-] \right|_{J_+ = J_- = 0}.
\end{equation}
Then non-vanishing contribution to the two-point function comes from the interaction Hamiltonian squared:
\begin{equation}
-\frac12 \left( \int dt H_\text{int}^+ - \int dt H_\text{int}^- \right)^2 = -\frac12 \int^t_{t_i} dt_1 \int^t_{t_i}dt_2 \Big[ H_\text{int}^+H_\text{int}^+ + H_\text{int}^-H_\text{int}^- - H_\text{int}^+H_\text{int}^- - H_\text{int}^-H_\text{int}^+ \Big] \, ,
\end{equation}
hence we come up with
\begin{align}
\left\langle \Omega \left| \Psi(t,\mathbi{x})\Psi(t,\mathbi{y}) \right| \Omega \right\rangle & = \left\langle \Omega \left| \Psi(t,\mathbi{x})\Psi(t,\mathbi{y}) \right| \Omega \right\rangle_\text{free} + \int_{t_i}^t dt_1 \int_{t_i}^t dt_2 \left\langle \Omega \left| H_\text{int}^+(t_1) \Psi(t,\mathbi{x})\Psi(t,\mathbi{y}) H_\text{int}^-(t_2) \right| \Omega \right\rangle 
\nonumber\\
& \quad - \int_{t_i}^t dt_1 \int_{t_i}^t dt_2 \left\langle \Omega \left| H_\text{int}^+(t_1) H_\text{int}^+(t_2) \Psi(t,\mathbi{x})\Psi(t,\mathbi{y}) \right| \Omega \right\rangle 
\nonumber\\
& \quad - \int_{t_i}^t dt_1 \int_{t_i}^t dt_2 \left\langle \Omega \left| \Psi(t,\mathbi{x})\Psi(t,\mathbi{y}) H_\text{int}^-(t_1) H_\text{int}^-(t_2) \right| \Omega \right\rangle + \cdots \, ,
\end{align}
where $\left\langle \Omega \left| \Psi(t,\mathbi{x})\Psi(t,\mathbi{y}) \right| \Omega \right\rangle_\text{free}$ denotes the two-point correlation function coming from the free quadratic action for $\Psi$, \eqref{eq:quadratic-free}. Since the interaction Hamiltonian contains two field contents, the correction terms now involve two propagators for $\Psi$ and one for the coupled field $\Phi$. Thus, in terms of the propagators we can write as
\begin{align}
\label{eq:general2point}
& \left\langle \Omega \left| \Psi(t,\mathbi{x})\Psi(t,\mathbi{y}) \right| \Omega \right\rangle 
\nonumber\\
= & i\Delta^\Psi_{+-}
-\int^t_{t_i} \bar{N}dt_1d^3x_1 \int^t_{t_i}\bar{N}dt_2d^3x_2 (a^3c)(t_1)(a^3c)(t_2) \calO_1^{(\Psi)} i\Delta^\Psi_{+-}(t, t_1) \calO_1^{(\Psi)} i\Delta^\Psi_{-+}(t,t_2) \calO_1^{(\Phi)}\calO_2^{(\Phi)} i\Delta^\Phi_{+-}(t_1, t_2)
\nonumber\\
& - \int^t_{t_i}\bar{N} dt_1 d^3x_1 \int^{t_1}_{t_i}\bar{N}dt_2 d^3x_2 (a^3c)(t_1)(a^3c)(t_2)
\left[ \calO_1^{(\Psi)} i\Delta^\Psi_{+-}(t_1, t) \calO_2^{(\Psi)} i\Delta^\Psi_{+-}(t, t_2) \calO_1^{(\Phi)}\calO_2^{(\Phi)} i\Delta^\Phi_{+-}(t_1, t_2) \right.
\nonumber\\
& \hspace{18em} \left. + \calO_1^{(\Psi)} i\Delta^\Psi_{-+}(t, t_1) \calO_2^{(\Psi)} i\Delta^\Psi_{-+}(t_2, t) \calO_1^{(\Phi)}\calO_2^{(\Phi)} i\Delta^\Phi_{+-}(t_2, t_1) \right] \, .
\end{align}
This is most general expression for the two-point correlation function of $\Psi$ including the leading corrections due to the interaction with $\Phi$. The corresponding power spectrum can be found by taking the Fourier transformation of $\left\langle \Omega \left| \Psi(t,\mathbi{x})\Psi(t,\mathbi{y}) \right| \Omega \right\rangle$.

For definiteness, let us consider explicitly the corrections to the power spectrum of $\calR$ due to the interaction with $\varphi_\bot^a$ in the simplest two-field case. The quadratic mixing term is rewritten as 
\begin{equation}
\label{Eq:quadmix}
S_\text{int} = \int d^3x \bar{N}dt \, a^3 \frac{2}{H} V_a\varphi_\bot^a \dot\calR = \int d^3x \bar{N}dt \, a^3 \sqrt{8\epsilon}\mpl\dot\theta \varphi\dot\calR \equiv \int d^3x \bar{N}dt \, a^3  c(t) \varphi \dot\calR \, .
\end{equation}
Comparing with \eqref{eq:general2point}, $c(t) = \sqrt{8\epsilon}\mpl\dot\theta$, $\calO^{(\calR)} = \partial_t$ and $\calO^{(\varphi)} = 1$. Then the two-point correlation of $\calR$ is given by
\begin{align}
& \left\langle \Omega \left| \calR(t,\mathbi{x})\calR(t,\mathbi{y}) \right| \Omega \right\rangle 
\nonumber\\
= & i\Delta^\calR_{+-}
-\int^t_{t_i} \bar{N}dt_1d^3x_1 \int^t_{t_i}\bar{N}dt_2d^3x_2 (a^3c)(t_1)(a^3c)(t_2) \partial_{t_1} i\Delta^\calR_{+-}(t, t_1) \partial_{t_2} i\Delta^\calR_{-+}(t,t_2) i\Delta^\varphi_{+-}(t_1, t_2)
\nonumber\\
& - \int^t_{t_i}\bar{N} dt_1 d^3x_1 \int^{t_1}_{t_i}\bar{N}dt_2 d^3x_2 (a^3c)(t_1)(a^3c)(t_2)
\left[ \partial_{t_1} i\Delta^\calR_{+-}(t_1, t) \partial_{t_2} i\Delta^\calR_{+-}(t, t_2) i\Delta^\varphi_{+-}(t_1, t_2) \right.
\nonumber\\
& \hspace{18em} \left. + \partial_{t_1} i\Delta^\calR_{-+}(t, t_1) \partial_{t_2} i\Delta^\calR_{-+}(t_2, t) i\Delta^\varphi_{+-}(t_2, t_1) \right] \, .
\end{align}
The power spectrum is obtained by performing the Fourier transformation of the above result. In the limit of constant $c(t)$, using the mode function solution \eqref{Eq:massivesol} for $\varphi$ for which $s = 1$, we find
\begin{equation}
\calP_\calR = \left( \frac{H}{2\pi} \right)^2 \frac{1}{2\epsilon\mpl^2} \left( 1 + \frac{4c^2C}{\epsilon\mpl^2H^2} \right) \, ,
\end{equation}
where~\cite{Chen:2009zp}
\begin{equation}
C \equiv \frac{\pi}{4} \Re \left\{ \int^\infty_0 dx_1 \int^\infty_{x_1} dx_2 \left[ x_1^{-1/2}H_\nu^{(1)}(x_1)e^{ix_1}x_2^{-1/2}H_\nu^{(2)}(x_2)e^{-ix_2} - x_1^{-1/2}H_\nu^{(1)}(x_1)e^{-ix_1}x_2^{-1/2}H_\nu^{(2)}(x_2)e^{-ix_2} \right] \right\} \, .
\end{equation}
Especially, if $\varphi$ is very light so that $\nu \to 3/2$, we find\footnote{Note that this result is obtained by using the asymptotic form of the exponential integral function ${\rm Ei}(x\to0)$ to perform the outermost integral. Thus our analytic estimate \eqref{eq:coeff-C} is not precisely the exact result, but the error is very likely to be $\calO(\gamma^2)$.}
\begin{equation}
\label{eq:coeff-C}
C = \frac{1}{2} (N_k - \gamma - \log2 + 2)^2 + \frac{\pi^2}{8} - \frac{\gamma^2}{2} - \frac{3}{4} \, ,
\end{equation}
where $\gamma \approx 0.577216$ is the Euler-Mascheroni constant and $N_k = -\log(-k\tau_e)$ is the number of $e$-folds elapsed between the moment of horizon crossing for the mode of our interest and the end of inflation $\tau_e$~\cite{cutoff}. For the field fluctuations $\varphi^a$ in the flat gauge, we can find similar results.

\section{Higher-order correlation functions}
\label{Sec:higherorder}
\setcounter{equation}{0}

\subsection{Cubic action}
\label{subsec:cubic}

Having discussed two-point correlation functions, now we consider higher order correlation function. They are necessary to investigate the properties of the cosmological perturbations and the physics behind them beyond linear perturbation theory. The first higher order correlation function is the three-point function, or its Fourier transform -- the bispectrum, which is the leading probe of non-Gaussianity. To compute three-point correlation functions, we need the cubic order action in perturbations. The cubic terms are obtained from the interaction Lagrangian $S_\text{int}$ in \eqref{Eq:Lagrangian}.

In the flat gauge, simply $\tilde\varphi^a = \varphi^a$ and the scalar cubic action becomes 
\begin{align}
S_3 & = \int d^3x\bar{N}dt a^3 \left[ -\frac{V_{abc}}{3!}\varphi^a\varphi^b\varphi^c \right.
\nonumber\\
& \hspace{6em} - \frac{\delta_{ab}\dot{\phi}_0^a\varphi^b}{4\mpl^2H} \left\{ \frac{1}{\mpl^2} \left[ \delta_{ij}\delta_{cd}\dot{\phi}_0^c\varphi^d - \frac{V}{4H^2} \left( \delta_{ij} - \frac{\partial_i\partial_j}{\nabla^2} \right) \left( \frac{2H}{V} \left( \delta_{cd}\dot{\phi}_0^c\dot{\varphi}^d + V_c\varphi^c \right) +\frac{2\delta_{cd}\dot{\phi}_0^c\varphi^d}{\mpl^2} \right) \right]^2 \right.
\nonumber\\
& \hspace{11em} - \frac{1}{2\mpl^2} \left[ 3\delta_{cd}\dot{\phi}_0^c\varphi^d - \frac{V}{2H^2} \left( \frac{2H}{V} \left( \delta_{cd}\dot{\phi}_0^c\dot{\varphi}^d + V_c\varphi^c \right) + \frac{2\delta_{cd}\dot{\phi}_0^c\varphi^d}{\mpl^2} \right) \right]^2
\nonumber\\
& \hspace{11em} \left. + \delta_{cd} \left( \frac{\delta_{ef}\dot{\phi}_0^e\varphi^f}{2\mpl^2H}\dot{\phi}_0^c - \dot{\varphi}^c \right) \left( \frac{\delta_{gh}\dot{\phi}_0^g\varphi^h}{2\mpl^2H}\dot{\phi}_0^d - \dot{\varphi}^d \right) 
 + \frac{1}{a^2}\delta_{cd}\partial^i\varphi^c\partial_i\varphi^d + V_{cd}\varphi^c\varphi^d \right\}
\nonumber\\
& \hspace{6em} \left. - \frac{V}{4\mpl^2H^2} \left( \frac{\delta_{ab}\dot{\phi}_0^a\varphi^b}{2\mpl^2H}\delta_{ef}\dot{\phi}_0^e\partial_i\varphi^f-\delta_{ab}\dot{\varphi}^a\partial_i\varphi^b \right) \frac{\partial^i}{\nabla^2} \left( \frac{2H}{V} \left( \delta_{cd}\dot{\phi}_0^c\dot{\varphi}^d + V_c\varphi^c \right) + \frac{2\delta_{cd}\dot{\phi}_0^c\varphi^d}{\mpl^2} \right) \right] \, .
\end{align}
Collecting the leading order terms, we have the standard terms~\cite{Maldacena:2002vr,multi-fieldfluc} extended to multi-field case:
\begin{equation}
S_3 \supset \int d^3x \bar{N}dt a^3 \frac{-\delta_{ab}\delta_{cd}}{4\mpl^2H} \left[ \dot\phi_0^a\varphi^b \left( \dot\varphi^c\dot\varphi^d + \frac{1}{a^2}\partial^i\varphi^c\partial_i\varphi^d \right) - 2\dot\phi_0^a \left( \frac{\partial_i}{\nabla^2}\dot\varphi^b \right) \dot\varphi^c\partial^i\varphi^d \right] \, ,
\end{equation}
from which we can find the leading bispectrum of the field fluctuations, evaluated at the moment of the horizon crossing~\cite{multi-fieldfluc}. The bispectrum of the curvature perturbation can be subsequently calculated by implementing the $\delta{N}$ formalism~\cite{deltaN}.

Meanwhile, in the comoving gauge, the mixing between the Goldstone mode $\pi$, which is related to the curvature perturbation $\calR$ by \eqref{eq:pi->R}, and the orthogonal  modes $\varphi_\bot^a$ becomes explicit. The action cubic in $\calR$ receives contributions also from quadratic terms in writing $\pi$ in terms of $\calR$. We find, to leading order,
\dis{
S_{\calR\calR\calR} = \int d^3x \bar{N}dt a^3\mpl^2 \bigg[ & \left (-\epsilon^2+2\epsilon\delta \right)\calR\dot\calR^2 - 2\epsilon^2 \dot\calR\partial^i\calR\frac{\partial_i}{\nabla^2}\dot\calR + \left( 3\epsilon^2-2\epsilon\delta \right) \calR \left( \frac{\partial_i\calR}{a} \right)^2
\\
& \left. + \frac14 \left( \frac{\epsilon V_{ab}\dot{\phi}_0^a \dot{\phi}_0^b}{H^2} + \frac{V_{abc}\dot{\phi}_0^a \dot{\phi}_0^b \dot{\phi}_0^c}{3H^3} + \frac{\delta_{ab}\ddot{\phi}_0^a\ddot{\phi}_0^b}{H^2} \right) \calR^3 \right] + {\cal O}(\epsilon^3) \, .
\label{Eq:cubic}
}
On the other hand, the $\pi$-$\varphi_\bot$ mixing term in the quadratic action gives the cubic order contributions as 
\dis{
S_\text{quadratic mixing} = \int d^3x \bar{N}dt a^3 \frac{V_a\varphi_\bot^a}{H} \bigg\{ & 
(3\epsilon-\delta)\calR\dot{\calR} + \frac{1}{H} \left( \calR\ddot{\calR}+\dot{\calR}^2 \right)
- \frac{1-\epsilon}{4a^2H} \left[ (\partial_i\calR)^2 - \frac{\partial^i\partial^j}{\nabla^2}\partial_i\calR\partial_j\calR \right]
\\
& + \frac{1}{2a^2H^2} \left( \partial^i\calR\partial_i\dot{\calR} - \frac{\partial^i\partial^i}{\nabla^2}\partial_i\calR\partial_j\dot{\calR} \right)
\\
& \left. + \frac{\epsilon}{H} \left(1-\frac{\partial^i\partial^j}{\nabla^2}\right) \left( \partial_i\dot{\calR}\frac{\partial_j}{\nabla^2}\calR + \partial_i\calR\frac{\partial_j}{\nabla^2}\dot{\calR} \right) \right\} + {\cal O}(\epsilon^2) \, . 
} 
Finally, the cubic action containing $\varphi_\bot^3$, $\calR\varphi_\bot^2$ and $\calR^2\varphi_\bot$ reads
\dis{
S_\text{cubic mixing} = \int d^3x \bar{N}dta^3 \bigg\{ & -\frac{V_{abc}}{6}\varphi_\bot^a \varphi_\bot^b \varphi_\bot^c -\frac{V_{abc}}{6 H^2} \left( \varphi_\bot^a \dot{\phi}_0^b \dot{\phi}_0^c+ \dot{\phi}_0^a\varphi_\bot^b  \dot{\phi}_0^c+ \dot{\phi}_0^a \dot{\phi}_0^b\varphi_\bot^c \right) \calR^2
\\
& + \frac{V_{abc}}{6H} \left( \varphi_\bot^a\varphi_\bot^b\dot{\phi}_0^c + \varphi_\bot^a\dot{\phi}_0^b\varphi_\bot^c + \dot{\phi}_0^a\varphi_\bot^b\varphi_\bot^c \right) \calR
\\
+ & \epsilon\calR \left[
- \frac{2}{\mpl^2H^2} \left( V_a\varphi_\bot^a \right)^2
+ \frac{2}{\mpl^2 H^2} \left( \frac{\partial_i\partial_i}{\nabla^2} \left( V_a \varphi_\bot^a \right) \right)^2 \right.
\\
& \quad\quad - 4\epsilon\calR (V_a \varphi_\bot^a)
+ \frac{2\epsilon}{H}\dot\calR(V_a \varphi_\bot^a)
- \frac{2\epsilon}{H} \left( \frac{\partial_i \partial_j}{\nabla^2}\dot\calR \right) \frac{\partial_i \partial_j}{\nabla^2}(V_a \varphi_\bot^a)
\\
& \quad\quad - \frac{1}{2H}\dot\calR \left( V_a\varphi_\bot^a \right) + \frac{1}{2H}\calR \left( 3HV_a\varphi_\bot^a+V_a\dot{\varphi}_\bot^a \right)
\\
& \quad\quad \left. + \frac12\delta_{ab} \left( \dot{\varphi}_\bot^a \dot{\varphi}_\bot^b-\partial_i \varphi_\bot^a \partial_i \varphi_\bot^b \right) - \frac12 M_{ab}^2 \left( \varphi_\bot^a \varphi_\bot^b - \frac{2\dot\phi^a_0}{H}\calR\varphi_\bot^b \right) \right]
\\
+ & \left[ \delta_{ab}\partial_i \varphi_\bot^a \dot{\varphi}_\bot^b - \frac{1}{H} \left( \partial^i \calR V_a \varphi_\bot^a + \calR V_a \partial^i \varphi^a \right) \right] \frac{\partial_i}{\nabla^2} \left( \epsilon\dot\calR-\frac{V_a\varphi_\bot^a}{\mpl^2H} \right)
\\
+ & \left. \frac{2}{H} \left( \epsilon \dot\calR\partial^i\calR - \delta H \calR \partial^i \calR \right)  \frac{\partial_i}{\nabla^2}(V_a\varphi_\bot^a) \right\} \, .
}

\subsection{Three-point function}

With the cubic order action found in the previous section, now we can straightly compute the three-point correlation functions. The essential steps are more or less the same as before. Formally, the three-point function of, say, $\calR$ is obtained by calculating the functional derivatives of the generating functional:
\dis{
\langle \Omega | \calR(t, \mathbi{x}) \calR(t, \mathbi{y}) \calR (t, \mathbi{z}) |\Omega \rangle = \left. \frac{\delta}{i\delta J_+(x)} \frac{\delta}{i\delta J_+(y)} \frac{\delta}{i\delta J_+(z)}  Z[J_+, J_-] \right|_{J_+=J_-=0} \, .
}
From \eqref{Eq:cubic}, the contribution from $S_{\calR\calR\calR}$ is given by
\dis{
& \langle \Omega | \calR(t, \mathbi{x}) \calR(t, \mathbi{y}) \calR (t, \mathbi{z}) |\Omega \rangle 
\\
= & \int^t_{-\infty}d^3w \bar{N}dt'a^3\mpl^2
\bigg[ \left(-\epsilon^2+2\epsilon\delta\right) \Delta_{+-}^\calR(t',\mathbi{w};t,\mathbi{x})\partial_{t'}\Delta_{+-}^\calR(t',\mathbi{w};t,\mathbi{y})\partial_{t'}\Delta_{+-}^\calR(t',\mathbi{w};t,\mathbi{z}) 
\\
& - 2\epsilon^2\partial_{t'}\Delta_{+-}^\calR(t',\mathbi{w};t,\mathbi{x})\partial^w_i\Delta_{+-}^\calR(t',\mathbi{w};t,\mathbi{y})\frac{\partial^w_i}{\nabla_w^2}\Delta_{+-}^\calR(t',\mathbi{w};t,\mathbi{z})
\\
& + \left(3\epsilon^2-2\epsilon \delta\right)\Delta_{+-}^\calR(t',\mathbi{w};t,\mathbi{x})\frac{\partial^w_i}{a}\Delta_{+-}^\calR(t',\mathbi{w};t,\mathbi{y})\frac{\partial^w_i}{a}\Delta_{+-}^\calR(t',\mathbi{w};t,\mathbi{z})
\\
& + \frac14 \left( \frac{\epsilon V_{ab}\dot{\phi}_0^a \dot{\phi}_0^b}{H^2} + \frac{V_{abc}\dot{\phi}_0^a \dot{\phi}_0^b \dot{\phi}_0^c}{3H^3} + \frac{\delta_{ab}\ddot{\phi}_0^a\ddot{\phi}_0^b}{H^2} \right)
\Delta_{+-}^{\cal R}(t',\mathbi{w};t,\mathbi{x})\Delta_{+-}^{\cal R}(t',\mathbi{w};t,\mathbi{y})\Delta_{+-}^{\cal R}(t',\mathbi{w};t,\mathbi{z})
+({\rm permutations})
\\
& - \left( -\epsilon^2+2\epsilon\delta \right) \Delta_{-+}^\calR(t',\mathbi{w};t,\mathbi{x})\partial_{t'}\Delta_{-+}^\calR(t',\mathbi{w};t,\mathbi{y})\partial_{t'}\Delta_{-+}^\calR(t',\mathbi{w};t,\mathbi{z})
\\
& + 2\epsilon^2\partial_{t'}\Delta_{-+}^\calR(t',\mathbi{w};t,\mathbi{x})\partial^w_i\Delta_{-+}^\calR(t',\mathbi{w};t,\mathbi{y})\frac{\partial^w_i}{\nabla_w^2}\Delta_{-+}^\calR(t',\mathbi{w};t,\mathbi{z})
\\
& - \left(3\epsilon^2-2\epsilon \delta\right)\Delta_{-+}^\calR(t',\mathbi{w};t,\mathbi{x})\frac{\partial^w_i}{a}\Delta_{-+}^\calR(t',\mathbi{w};t,\mathbi{y})\frac{\partial^w_i}{a}\Delta_{-+}^\calR(t',\mathbi{w};t,\mathbi{z})
\\
& - \frac14 \left( \frac{\epsilon V_{ab}\dot{\phi}_0^a \dot{\phi}_0^b}{H^2} + \frac{V_{abc}\dot{\phi}_0^a \dot{\phi}_0^b \dot{\phi}_0^c}{3H^3} + \frac{\delta_{ab}\ddot{\phi}_0^a\ddot{\phi}_0^b}{H^2} \right)
\Delta_{-+}^\calR(t',\mathbi{w};t,\mathbi{x})\Delta_{-+}^\calR(t',\mathbi{w};t,\mathbi{y})\Delta_{-+}^\calR(t',\mathbi{w};t,\mathbi{z})
-({\rm permutations}) \bigg] \, .
}
This is a good approximation for the three-point function of $\calR$ when the other fields $\varphi_\bot$ are much heavier than $\sqrt{\epsilon}H$ so that they decouple completely.

If some of the $\varphi_\bot$'s are not heavy, which is indeed  the case for ``multi-field'' inflation, the contribution of $\varphi_\bot$  to the three-point function of ${\cal R}$ becomes manifest. In Section~\ref{subsec:cubic}, we found that the cubic interaction contains various $\varphi_\bot \calR^2$, $\varphi_\bot^2\calR$ and $\varphi_\bot^3$ vertices. Through the quadratic mixing term \eqref{Eq:quadmix}, $\varphi_\bot$ in the cubic action is converted to $\calR$ to contribute to the three-point correlation function of $\calR$ for the simplest two-field case that we have discussed in Section~\ref{subsec:powerspec}, in the presence of the cubic interaction ${\cal O} \varphi_\bot {\cal R}^2$, with $\calO$ denoting collectively the coefficients and derivative operators acting on $\varphi_\bot\calR^2$ terms, the contribution to $\langle \Omega | \calR(t, \mathbi{x}) \calR(t, \mathbi{y}) \calR (t, \mathbi{z}) |\Omega \rangle$ is given by
\dis{
& \langle \Omega | \calR(t, \mathbi{x}) \calR(t, \mathbi{y}) \calR (t, \mathbi{z}) |\Omega \rangle 
\\
\supset & \int^t_{-\infty} d^3w \bar{N}dt'{\cal O} \bigg[ \Delta_{+-}^{\cal R}(t',\mathbi{w};t,\mathbi{x})\Delta_{+-}^{\cal R}(t',\mathbi{w};t,\mathbi{y})
\\
& \qquad \times \int^{t'}_{-\infty}d^3x' \bar{N}dt'' \left( \frac{a^3}{H}V_a \right)(t'') \Delta_{+-}^{\varphi_\bot}(t',\mathbi{w};t'',\mathbi{x}')\partial_{t''}\Delta_{+-}^{\cal R}(t'',\mathbi{x}';t,\mathbi{x}) +({\rm permutations})
\\
& \quad -\Delta_{-+}^{\cal R}(t',\mathbi{w};t,\mathbi{x})\Delta_{-+}^{\cal R}(t',\mathbi{w};t,\mathbi{y})
\\
& \qquad \times\int^{t'}_{-\infty}d^3x' \bar{N}dt'' \left(\frac{a^3}{H}V_a \right)(t'') \Delta_{-+}^{\varphi_\bot}(t',\mathbi{w};t'',\mathbi{x}')\partial_{t''}\Delta_{-+}^{\cal R}(t'',\mathbi{x}';t,\mathbi{x}) +({\rm permutations}) \bigg] \, .
}

\section{Conclusions}
\label{Sec:conclusion}

In this article, we have developed the path integral formalism for inflationary cosmology in the presence of multiple scalar fields. From a theoretical point of view, the path integral formalism is advantageous in treating a quantum theory with gauge symmetry. In the presence of gravity, the ADM formalism makes diffeomorphism as a gauge symmetry evident. This motivates us to apply the path integral approach to inflation. Since an FRW spacetime which is not exactly de Sitter breaks time translational invariance spontaneously, we expect the associated Goldstone mode to be extracted as a combination of scalar fields, which can be treated as massless at energies above $\sqrt{\epsilon}H$. The $n$-point correlation functions  of the Goldstone mode contain effects of the other degrees of freedom. The deviation from single field inflation in the presence of extra light degrees of freedom can be tested in various observations. Whereas our consistency checks have been made at tree-level, the path integral formalism is expected to be useful when we consider the loop level as it provides a consistent and systematic way to describe the quantum behavior of scalar fields in the presence of gravity.

The main purpose of the present work is to lay down the path integral framework for multi-field inflation. We illustrated our approach with the simplest scenarios. It is straightforward to generalize our analysis to compute the non-Gaussianities of $P(X,\phi)$ theories of inflation~\cite{Chen:2006nt},  such as the multi-field generalization~\cite{Huang:2007hh} of DBI inflation~\cite{Alishahiha:2004eh}. It would also be interesting to examine, using the path integral approach, consistency conditions among correlation functions for multi-field inflation. We plan to return to these issues in future work.

\subsection*{Acknowledgements}

JG thanks the Institute for Advanced Study, Hong Kong University of Science and Technology for hospitality during the Gordon Research Conference ``String Theory \& Cosmology: New Ideas Meet New Experimental Data'', where this work was initiated.
JG acknowledges support from the Korea Ministry of Education, Science and Technology, Gyeongsangbuk-Do and Pohang City for Independent Junior Research Groups at the Asia Pacific Center for Theoretical Physics. 
JG is also supported in part by a Starting Grant through the Basic Science Research Program of the National Research Foundation of Korea (2013R1A1A1006701) and by TJ Park Science Fellowship of POSCO TJ Park Foundation.
MS is supported by IBS under the project code, IBS-R018-D1.
GS is supported by part by the DOE grant DE-FG-02-95ER40896 and the HKRGC grants HUKST4/CRF/13G, 604231 and 16304414.

\appendix

\renewcommand{\theequation}{\Alph{section}.\arabic{equation}}

\section{Completion of the action for cosmological perturbations}
\label{app:lagrangian}
\setcounter{equation}{0}

Here, we list each term appearing in the full perturbation action in \eqref{Eq:Lagrangian0} and \eqref{Eq:Lagrangian} in detail. The free Hamiltonian and constraints are given by
\begin{align}
\calH_\text{free} & = 4\bar{N}a^3H^2\frac{\mpl^2}{2} \left[ \frac12 \pi^{ij}A_{ijkl}\pi^{kl} + \pi^{ij} \left( 2h_{ij} - \frac12\delta_{ij}h \right) \right] + \bar{N}\frac{\delta^{ab}}{2a^3} \left( \pi_a\pi_b - h P_a\pi_b \right)
\nonumber\\
& \quad +\bar{N}\frac{a}{2}\delta_{ab}\partial_i \varphi^a \partial_i \varphi^b + {\bar N}\frac{a^3}{2} \left( V_{ab}\varphi^a\varphi^b+h V_a\varphi^a \right)
\nonumber\\
& \quad + {\bar N} \left( \frac{5\mpl^2}{4} a^3H^2 + \frac{\delta^{ab}}{8a^3}P_aP_b \right) h^{ij}h_{ij} - {\bar N} \left( \frac{3\mpl^2}{8} a^3 H^2 - \frac{\delta^{ab}}{16 a^3} P_aP_b \right) h^2
\nonumber\\
& \quad + {\bar N}a\frac{\mpl^2}{2} \left( \frac14 h\nabla^2 h - \frac12h h^{ij}_{,ij} + \frac12 h^{ij}\partial^l\partial_ih_{jl} - \frac14 h^{ij}\nabla^2 h_{ij} \right) + {\bar N}a^3\frac{V}{4} \left( \frac{h^2}{2} - h^{ij}h_{ij} \right) \, ,
\\
C^{0}_1 & = a^3h \left( \frac{\delta^{ab}P_aP_b}{4a^6} + \frac{\mpl^2}{2}H^2 \right) + a\frac{\mpl^2}{2} \left( h^{ij}_{,ij}-\nabla^2 h \right) - a^3 V_a\varphi^a - \frac{\delta^{ab}}{a^3}P_a\pi_b + 2\mpl^2a^3H^2\pi - \frac{a^3}{2}hV \, ,
\label{Eq:C0(1)}
\\
C^i_1 & = -\frac{1}{a^2}P_a \partial^i\varphi^a - 2aH\mpl^2 \left( \partial_j\pi^{ij} + \partial_j h^{ij} - \frac12 \partial^i h \right) \, .
\end{align}
Meanwhile, the interaction Hamiltonian and constraints are
\begin{align}
\calH_\text{int} & = 4{\bar N}a^3 H^2\frac{\mpl^2}{2} \left\{ -\frac32\left({\tilde \gamma}^{-1/2}\right)^{\geq 3} - \left({\tilde \gamma}^{-1/2}\right)^{\geq 2}h + \left({\tilde \gamma}^{-1/2}\right)^{\geq 1}\frac12 h^{ij}A_{ijkl}h^{kl} \right.
\nonumber\\
& \qquad\qquad\qquad + \pi^{ij} \left[ -\left({\tilde \gamma}^{-1/2}\right)^{\geq 2}\delta_{ij} + \left({\tilde \gamma}^{-1/2}\right)^{\geq 1} \left(h_{ij}-\delta_{ij}h\right) + \left({\tilde \gamma}^{-1/2}\right) \left(2h_i{}^kh_{jk}-hh_{ij}\right) \right]
\nonumber\\
& \qquad\qquad\qquad \left. + \pi^{ij} \left[ \left({\tilde \gamma}^{-1/2}\right)^{\geq 1}\frac12 A_{ijkl} + \left({\tilde \gamma}^{-1/2}\right) \left( 2h_{jl}\delta_{ik} - \delta_{ij}h_{kl} + h_{ik}h_{jl} - \frac12h_{ij}h_{kl} \right) \right]\pi^{kl} \right\}
\nonumber\\
& \quad + {\bar N}a^3 \left[ \left({\tilde \gamma}^{-1/2}\right)^{\geq 2}\sum_{n=1}^\infty \frac{V^{(n)}}{n!}\varphi^n + \frac{h}{2}\sum_{n=1}^\infty \frac{V^{(n)}}{n!}\varphi^n + \sum_{n=3}^\infty \frac{V^{(n)}}{n!}\varphi^n + \left({\tilde \gamma}^{-1/2}\right)^{\geq 3}V \right]
\nonumber\\
& \quad + {\bar N}\frac{\delta^{ab}}{2a^3} \left[ \left({\tilde \gamma}^{-1/2}\right)^{\geq 1}\pi_a\pi_b + 2\left({\tilde \gamma}^{-1/2}\right)^{\geq 2} P_a\pi_b + \left({\tilde \gamma}^{-1/2}\right)^{\geq 3}P_aP_b \right]
\nonumber\\
& \quad + \frac{a{\bar N}}{2} \left[ \left({\tilde \gamma}^{1/2}\right)^{\geq 1}\delta^{ij} + \left({\tilde \gamma}^{ij}\right)^{\geq 1} + \left({\tilde \gamma}^{1/2}\right)^{\geq 1}\left({\tilde \gamma}^{ij}\right)^{\geq 1} \right] \delta_{ab}\partial_i\varphi^a\partial_j\varphi^b
\nonumber\\
& \quad - a{\bar N} \left\{ \left({\tilde \gamma}^{1/2}\right)^{\geq 1} \left[ \left({\tilde \gamma}^{ij}\right)^{\geq 1}\delta^{kl} + \delta^{ij}\left({\tilde \gamma}^{kl}\right)^{\geq 1} \right] + {\tilde \gamma}^{1/2}\left({\tilde \gamma}^{ij}\right)^{\geq 1}\left({\tilde \gamma}^{kl}\right)^{\geq 1} \right\} \frac{\mpl^2}{2} \left( \partial_i\partial_k h_{jl} - \partial_i\partial_j h_{kl} \right)
\nonumber\\
& \quad - a{\bar N}\frac{\mpl^2}{2} \left( {\tilde \gamma}^{1/2}{\tilde \gamma}^{ij}{\tilde \gamma}^{km}{\tilde \gamma}^{ln}-\delta^{ij}\delta^{km}\delta^{ln} \right) 
\nonumber\\
& \qquad\qquad \times \left( -h_{mn,k}h_{jl,i}-\frac14 h_{jm,l}h_{in,k}-\frac14 h_{ij,n}h_{km,l}+h_{ij,l}h_{mn,k}+\frac34h_{kl,i}h_{mn,j} \right) \, ,
\\
C^0_{\geq 2} & = -\frac{1}{\bar N} \left( {\cal H}_\text{free}+{\cal H}_\text{int}-{\bar N}4a^3H^2 \frac{\mpl^2}{2}\pi^{ij}h_{ij} \right) \, ,
\\
C^i_{\geq 2} & = -\frac{1}{a^2}\left({\tilde \gamma}^{ij}\right)^{\geq 1} P_a \partial_j \varphi^a - \frac{1}{a^2}{\tilde \gamma}^{ij}\pi_a \partial_j\varphi^a - 4aH\frac{\mpl^2}{2} \left[ \left({\tilde \gamma}^{ij}\right)^{\geq 1} \left( {h_{jl,l}} - \frac12\partial_j h \right) + {\tilde \gamma}^{ik} \left( h_{kl,j}-\frac12 h_{jl,k} \right)\pi^{jl} \right] \, ,
\end{align}
where $A_{ijkl}\equiv \delta_{ik}\delta_{jl}+\delta_{il}\delta_{jk}-\delta_{ij}\delta_{kl}$.

\section{Auxiliary field terms}
\label{App:quadmom}
\setcounter{equation}{0}

In the quadratic action \eqref{Eq:quadratic}, the auxiliary field terms are separated from the dynamical fields after appropriate redefinitions. These are given as:
\begin{align}
({\rm auxiliary~field~terms}) = &\int d^4x \bar{N}a^3 \left\{ -\frac{1}{2a^6}\delta^{ab}\rho_a \rho_b- 2 \mpl^2 H^2\rho^{ij}\frac{A_{ijkl}}{2}\rho^{kl}-\frac{V}{{\bar N}^2}\tilde{n}^2 \right.
\nonumber\\
& \qquad\qquad\quad \left. + \frac{\mpl^2}{2a^4 {\bar N}^2} \left[ \left(\partial_{(i}\tilde{N}^{T}_{j)}\right)^2 + \frac{4H^2}{V}(\nabla^2 S)^2 \right] \right\} \, ,
\end{align}
where
\dis{
\rho_a & \equiv \pi_a+\frac12\Big[\Big(-h+\frac{2n}{\bar{N}}\Big)P_a-2a^3\delta_{ab}\dot{\varphi}^b\Big] \, ,
\\
\rho^{ij} & \equiv \pi^{ij}+\frac12(I^{ij}-\delta^{ij}I),
\\
I_{ij} & \equiv \frac{\dot{h}_{ij}}{2\bar{N}H}+2h_{ij}-\frac{h}{2}\delta_{ij}-\frac{n}{\bar{N}}\delta_{ij}-\frac{1}{\bar{N}a^2H}\partial_{(i}N_{j)} \, ,
\\
I & = \delta^{ij} I_{ij} \, ,
}
and 
\dis{
N_i & = \partial_i S+N_i^T \, ,
\\
\tilde{n} & = n - \frac{{\bar N}}{2a^3 V} \left[ a^3\mpl^2H\dot{h} - P_a \dot{\varphi}^a - a^3V_a\varphi^a + \frac{a\mpl^2}{2} \left( h^{ij}_{,ij}-\nabla^2 h \right) - \frac{2aH\mpl^2}{{\bar N}}\nabla^2 S \right] \, ,
\\
\partial_{(i}\tilde{N}^T_{j)} & = \partial_{(i}N^T_{j)}-\frac{a^2}{2}\bar{N}\partial_{(i}\dot{h}^T_{j)} \, ,
\\
\nabla^2 \tilde{S} & = \nabla^2 S+\frac{V}{12\mpl^2 H^2} \left( J-4a^2\bar{N}\nabla^2 \dot{H}_T \right) \, ,
\\
J_{ij} & = \bar{N}a^2 \left\{ -\dot{h}_{ij} + \delta_{ij} \left[ \left( 1-\frac{2\mpl^2H^2}{V} \right)\dot{h} - \frac{\mpl^2 H}{a^2V} \left( h^{ij}_{,ij}-\nabla^2 h \right) + \frac{2HP_a\dot{\varphi}^a}{a^3V} + \left( \frac{2P_a}{\mpl^2 a^3} + \frac{2HV_a}{V} \right) \varphi^a \right] \right\} \, ,
\\
J & = \delta^{ij}J_{ij} \, .
}

\section{Auxiliary fields and ghosts propagators}
\label{app:ghost}
\setcounter{equation}{0}

As discussed in Section~\ref{Sec:pathconstraints}, the dynamics of a constrained system is regulated by physical degrees of freedom and it is correctly described by inserting appropriate gauge fixing conditions and the Faddeev-Popov determinant. The Faddeev-Popov determinant is implemented in the Feynman rules by introducing ghosts $\bar{\eta}$ and $\eta$ as 
\begin{align}
K(t_f;t_i) & = \int {\cal D} \tilde{\varphi}^a {\cal D}h_{ij}^{TT} {\cal D}H_L {\cal D}H_T{\cal D} h_i^T {\cal D}\sigma \prod_\mu \delta (\psi_\mu) \left| {\rm det} \left\{ \psi_\mu, C_\nu \right\} \right| e^{iS}
\nonumber\\
&=\int {\cal D} \tilde{\varphi}^a {\cal D}h_{ij}^{TT} {\cal D}H_L {\cal D}H_T {\cal D} h_i^T {\cal D}\bar{\eta} {\cal D}\eta{\cal D}\sigma \prod_\mu \delta (\psi_\mu) \exp \left[ iS + i\bar\eta \left\{ \psi_\mu,C_\nu \right\} \eta \right] \, ,
\end{align}
where ${\cal D}\sigma \equiv {\cal D}\rho^{ij} {\cal D}\rho_a{\cal D} \tilde{n} {\cal D}\tilde{S} {\cal D}\tilde{N}_i^T$ denotes the integral measure for auxiliary fields, i.e. the fields without dynamics, resulting from redefinitions as listed in Appendix~\ref{App:quadmom}. These auxiliary fields are to be integrated out, and especially, if they are at most quadratic in the action, we just solve the equations of motion and then put their solutions back to the action. However, the integration is not always simple. For instance, the auxiliary fields in $\left| {\rm det} \left\{ \psi_\mu, C_\nu \right\} \right|$ may make the integration complicated. So, for a simple form of interaction Lagrangian, it is also allowed to include auxiliary fields in the Feynman rule with propagators \cite{Prokopec:2010be}:
\dis{
i\triangle^{\rho_a}_{\pm\pm} & = \mp \frac{a^3}{\bar{N}}i\delta^{(4)}(x-x') \, ,
\\
i \left( \triangle^\rho_{\pm\pm} \right)_{ijkl} & =\mp\frac{1}{8\mpl^2{\bar N}a^3H^2}(\delta_{ik}\delta_{jl}+\delta_{il}\delta_{jk}-2\delta_{ij}\delta_{kl})i\delta^{(4)}(x-x') \, ,
\\
i\triangle^{\tilde{n}}_{\pm \pm} & = \mp \frac{\bar{N}}{4 (3-\epsilon)a^3H^2}i\delta^{(4)}(x-x') \, ,
\\
i \left( \triangle^{\tilde{N}^T}_{\pm\pm} \right)_{ij} & = \mp \frac{2a\bar{N}}{\mpl^2}\nabla_x^{-2}P_{ij}i\delta^{(4)}(x-x') \, ,
\\
i\triangle^{\tilde S}_{\pm\pm} & = \pm \frac{a{\bar N}}{2\mpl^2}(3-\epsilon)\nabla_x^{-4}i\delta^{(4)}(x-x') \, ,
}
where $V=(3-\epsilon)\mpl^2H^2$ is used.

In order to obtain the ghost Feynman rule, we need  $\left.\left\{ \psi_\mu, C_\nu \right\} \right|_{\psi_\mu=0}$ calculated in Section~\ref{subsec:gaugefix}. Let $\Omega_{\mu\nu}$ be the free part of $\left.\left\{ \psi_\mu, C_\nu \right\} \right|_{\psi_\mu=0}$, i.e. the part which does not depend on the dynamical fields. Then the ghost propagators are given by 
\dis{
\Omega_{\mu\nu} (i\triangle^\eta)_{\pm\pm}(x, x') & = \pm i\delta_{\mu\nu}\delta^{(4)}(x-x') \, ,
\\
\Omega_{\mu\nu} (i\triangle^\eta)_{\pm\mp}(x, x') & = 0 \, .
}

\begin{itemize}

\item[1.] 
We first consider the flat gauge \eqref{eq:flatgauge}, From \eqref{Eq:hscalar1}, \eqref{Eq:hscalar2}, \eqref{Eq:hvector1} and \eqref{Eq:hvector2}, we have
\begin{align}
\left.\left\{ \psi_0, C_0 \right\}\right|_{\psi_\mu=0} & = -2H \left( \tilde{\gamma}^{-1/2} \right)^{TT} \left\{ 3+\rho+I-2h_{ij}^{TT}h_{ij}^{TT} -  \left[ \rho^{ij}-\frac12 \left( I^{ij}-\delta^{ij}I \right) \right]h_{ij}^{TT} \right.
\nonumber\\
& \hspace{6em} \left. - 2h_{ik}^{TT} \left[ \rho^{kl}-\frac12 \left( I^{kl}-\delta^{kl}I \right) \right]h_{li}^{TT} \right\}(t, \mathbi{y})\delta^{(3)}(\mathbi{x}-\mathbi{y}) \, ,
\\
\left.\left\{ \psi_0, C_i \right\}\right|_{\psi_\mu=0} & = -\frac{2}{a^2}\partial^x_i\delta^{(3)}(\mathbi{x}-\mathbi{y}) \, ,
\\
\left.\left\{ \psi_i, C_0 \right\}\right|_{\psi_\mu=0} & = -2H \left( \tilde{\gamma}^{-1/2} \right)^{TT} \left\{ -h_{ij}^{TT} - 2 \left[ \rho^{ij}-\frac12 \left( I^{ij}-\delta^{ij}I \right) - \frac{\delta_{ij}}{3}(\rho+I) \right] \right.
\nonumber\\
& \hspace{6em} + (\rho+I)h_{ij}^{TT} - 2 \left( h_{il}^{TT}h_{jl}^{TT}-\frac{\delta_{ij}}{3}h_{kl}^{TT}h_{kl}^{TT} \right)
\nonumber\\
& \hspace{6em} - 2 \left[ \left( \rho^{il} - \frac12 \left( I^{il}-\delta^{il}I \right) \right) h_{jl}^{TT} + \left( \rho^{jl} - \frac12 \left( I^{jl}-\delta^{jl}I \right) \right) h_{il}^{TT} \right.
\nonumber\\
& \hspace{8em} \left. - \frac23\delta_{ij} \left( \rho^{kl}-\frac12 \left( I^{kl}-\delta^{kl}I \right) \right)h_{kl}^{TT} \right]
\nonumber\\
& \hspace{6em} - 2 \left[ h_{il}^{TT}h_{jk}^{TT} \left( \rho^{kl} - \frac12 \left( I^{kl}-\delta^{kl}I \right) \right) - \frac{\delta_{ij}}{3}h_{km}^{TT}h_{lm}^{TT} \left( \rho^{kl}-\frac12 \left( I^{kl}-\delta^{kl}I \right) \right) \right]
\nonumber\\
& \hspace{6em} \left. + h_{ij}^{TT}h_{kl}^{TT} \left[ \rho^{kl} - \frac12 \left( I^{kl}-\delta^{kl}I \right) \right] \right\} (t, \mathbi{y})\delta^{(3)}( \mathbi{x}-\mathbi{y}),
\\
\left.\left\{ \psi_i, C_j \right\}\right|_{\psi_\mu=0} & = -\frac{1}{a^2} \left[ \delta_{ij}\nabla^2_x + \frac13\partial_i^x\partial_j^x - \left( \tilde{\gamma}^{jk} \right)^{TT} \left( h_{lk,i}^{TT}+h_{ik,l}^{TT}-h_{il,k}^{TT} \right) (t, \mathbi{y})\partial_l^x \right] \delta^{(3)}(\mathbi{x}-\mathbi{y}) \, ,
\end{align}
where the flat gauge conditions $H_L=0$, $H_T=0$ and $h_i^T=0$ are imposed here such that
\dis{
\left( \tilde{\gamma}^{ij} \right)^{TT} & = \delta^{ij} - h_{ij}^{TT} + h_{il}^{TT}h_{lj}^{TT} + \cdots \, ,
\\
\left( \tilde{\gamma}^{-1/2} \right)^{TT} & = 1 + \frac14 h_{ij}^{TT}h_{ij}^{TT} + \cdots \, .
}
Then, $\Omega_{\mu\nu}$ is given by, in the matrix form,
\dis{
\Omega=\left( 
\begin{array}{cc}
-6H & -\dfrac{2}{a^2}\partial_j^x
\\
0 & -\dfrac{1}{a^2} \left( \delta_{ij}\nabla_x^2 + \dfrac13\partial_i^x \partial_j^x \right)
\end{array}
\right) \, ,
}
from which the ghost propagators are given by
\dis{
(i\triangle^\eta_{\mu\nu})_{\pm\pm} = \left( 
\begin{array}{cc}
\mp \dfrac{1}{6H} & \pm \dfrac{1}{4H}\dfrac{\partial_j^x}{\nabla_x^2}
\\
0 & \mp a^2 \left( \delta_{ij} - \dfrac14 \dfrac{\partial_i^x \partial_j^x}{\nabla_x^2} \right) \dfrac{1}{\nabla_x^2}
\end{array}
\right)i\delta^{(4)}(x-x') \, . 
}

\item[2.] 
Next we consider the comoving gauge \eqref{eq:comgauge}. Likewise, from \eqref{Eq:phiscalar1}, \eqref{Eq:phiscalar2}, \eqref{Eq:hvector1} and \eqref{Eq:hvector2}, we have 
\begin{align}
\left.\left\{ \psi_0, C_0 \right\}\right|_{\psi_\mu=0} & 
= -\tilde{\gamma}^{-1/2} 2\epsilon\mpl^2 H^2 \delta^{(3)}(\mathbi{x}-\mathbi{y}) \, ,
\\
\left.\left\{ \psi_0, C_i \right\}\right|_{\psi_\mu=0} & =
0 \, ,
\\  
\left.\left\{ \psi_i, C_0 \right\}\right|_{\psi_\mu=0} & = -2H \left( \tilde{\gamma}^{-1/2} \right) \left\{ -h_{ij}^{TT} - 2 \left[ \rho^{ij} - \frac12 \left( I^{ij}-\delta^{ij}I \right) - \frac{\delta_{ij}}{3}(\rho+I) \right] \right.
\nonumber\\
& \hspace{6em} + \left( -\frac13h+\rho+I \right)h_{ij}^{TT} - 2 \left( h_{il}^{TT}h_{jl}^{TT}-\frac{\delta_{ij}}{3}h_{kl}^{TT}h_{kl}^{TT} \right)
\nonumber\\
& \hspace{6em}  - \frac43h \left[ \rho^{ij} - \frac12 \left( I^{ij}-\delta^{ij}I \right) - \frac{\delta_{ij}}{3}(\rho+I) \right]
\nonumber\\
& \hspace{6em}  - 2 \left[ \left( \rho^{il}-\frac12 \left( I^{il}-\delta^{il}I \right) \right) h_{jl}^{TT} + \left( \rho^{jl} - \frac12 \left( I^{jl}-\delta^{jl}I \right) \right) h_{il}^{TT} \right.
\nonumber\\
& \hspace{8em}  \left. - \frac23\delta_{ij} \left( \rho^{kl}-\frac12 \left( I^{kl}-\delta^{kl}I \right) \right) h_{kl}^{TT} \right]
\nonumber\\
& \hspace{6em}  - 2 \left[ \left( \frac{\delta_{il}}{3}h+h_{il}^{TT} \right) \left( \frac{\delta_{jk}}{3}h+h_{jk}^{TT} \right) \left( \rho^{kl}-\frac12 \left(I^{kl}-\delta^{kl}I \right) \right) \right.
\nonumber\\
& \hspace{8em}  \left. - \frac{\delta_{ij}}{3} \left( \frac{\delta_{km}}{3}h+h_{km}^{TT} \right) \left( \frac{\delta_{lm}}{3}h+h_{lm}^{TT} \right) \left( \rho^{kl}-\frac12 \left( I^{kl}-\delta^{kl}I \right) \right) \right]
\nonumber\\
& \hspace{6em}  \left. + h_{ij}^{TT} \left( \frac{\delta_{kl}}{3}h+h_{kl}^{TT} \right) \left[ \rho^{kl}-\frac12 \left( I^{kl}-\delta^{kl}I \right) \right] \right\} (t, \mathbi{y})\delta^{(3)}( \mathbi{x}-\mathbi{y}) \, ,
\\
\left.\left\{ \psi_i, C_j \right\}\right|_{\psi_\mu=0} & = -\frac{1}{a^2} \left[ \delta_{ij}\nabla^2_x + \frac13\partial_i^x\partial_j^x - \frac13\tilde{\gamma}^{jk} \left( h_{,i}\delta_{kl}+h_{,l}\delta_{ik}-h_{,k}\delta_{il} \right) \partial^x_l \right.
\nonumber\\
& \hspace{3.5em}  \left. - \tilde{\gamma}^{jk} \left( h_{lk,i}^{TT}+h_{ik,l}^{TT}-h_{il,k}^{TT} \right) (t, \mathbi{y})\partial_l^x-\frac19\tilde{\gamma}^{jk}h_{,k}(t, \mathbi{y})\partial_i^x \right] \delta^{(3)}(\mathbi{x}-\mathbi{y}) \, ,
\end{align}
where the comoving gauge conditions $\delta_{ab} \dot{\phi}_0^a \varphi^b=0$, $H_T=0$ and $h_i^T=0$ are imposed here. Then, $\Omega_{\mu\nu}$ in the matrix form is given by
\dis{
\Omega = \left( 
\begin{array}{cc}
-2\epsilon\mpl^2 H^2  & 0 \\
0 & -\dfrac{1}{a^2} \left( \delta_{ij}\nabla_x^2 + \dfrac13\partial_i^x \partial_j^x \right)
\end{array}
\right) \, .
}
From this, the ghost propagators are given by
\dis{
(i\triangle^\eta_{\mu\nu})_{\pm\pm} = \left( 
\begin{array}{cc}
\mp \dfrac{1}{2\epsilon\mpl^2H^2} & 0
\\
0 & \mp a^2 \left( \delta_{ij} - \dfrac14\dfrac{\partial_i^x \partial_j^x}{\nabla_x^2} \right) \dfrac{1}{\nabla_x^2}
\end{array}
\right)i\delta^{(4)}(x-x') \, . 
}

\end{itemize}

\end{document}